\documentclass[sigconf, nonacm]{acmart}

\usepackage{graphicx, amssymb, subfigure}
\usepackage{epstopdf}
\usepackage{algorithm}
\usepackage[noend]{algpseudocode}
\usepackage{geometry}
\usepackage{multicol}
\usepackage{arydshln}
\usepackage{amsthm}
\usepackage{subfiles}
\usepackage[section]{placeins}

\usepackage{amsmath}\allowdisplaybreaks
\usepackage{amsfonts,bm,bbm}
\usepackage{amssymb}

\def\eqref#1{equation~(\ref{#1})}

\def\1{\bf{1}}

\newcommand{\Norm}[1]{\left\| #1 \right\|}
\newcommand{\norm}[1]{\left\| #1 \right\|_2}

\def\va{{\bf{a}}}
\def\vb{{\bf{b}}}
\def\vc{{\bf{c}}}
\def\vd{{\bf{d}}}

\def\vu{{\bf{u}}}
\def\vv{{\bf{v}}}

\def\vx{{\bf{x}}}
\def\vy{{\bf{y}}}

\def\fZ{{\mathcal{Z}}}

\def\mA {{\bf A}}
\def\mB {{\bf B}}
\def\mC {{\bf C}}
\def\mD {{\bf D}}

\def\mI {{\bf I}}

\def\mK {{\bf K}}
\def\mL {{\bf L}}

\def\mP {{\bf P}}
\def\mQ {{\bf Q}}
\def\mR {{\bf R}}

\def\mU {{\bf U}}
\def\mV {{\bf V}}

\def\mX {{\bf X}}
\def\mY {{\bf Y}}

\def\mSigma {{\bf \Sigma}}

\usepackage{amsthm}
\theoremstyle{plain}
\newtheorem{thm}{Theorem}

\newtheorem{lem}{Lemma}

\makeatletter
\def\Ddots{\mathinner{\mkern1mu\raise\p@
\vbox{\kern7\p@\hbox{.}}\mkern2mu
\raise4\p@\hbox{.}\mkern2mu\raise7\p@\hbox{.}\mkern1mu}}
\makeatother

\makeatletter
\newcommand*{\rom}[1]{\expandafter\@slowromancap\romannumeral #1@}
\makeatother

\theoremstyle{definition}
\newtheorem{defn}{Definition}[section]
\newtheorem{corollary}{Corollary}[section]
\AtBeginDocument{%
  }

\begin{document}

\title{Optimal Approximate Matrix Multiplication over Sliding Windows}

\author{Ziqi Yao}
\affiliation{%
  \institution{East China Normal University}
  \country{Shanghai, China}
}
\email{51265902073@stu.ecnu.edu.cn}
\author{Mingsong Chen}
\affiliation{%
  \institution{East China Normal University}
  \country{Shanghai, China}
}
\email{mschen@sei.ecnu.edu.cn}
\author{Cheng Chen}
\authornote{Corresponding author}
\affiliation{%
  \institution{East China Normal University}
  \country{Shanghai, China}
}
\email{chchen@sei.ecnu.edu.cn}

\renewcommand{\shortauthors}{Ziqi Yao, Mingsong Chen, Cheng Chen}

\begin{abstract}
We explore the problem of approximate matrix multiplication (AMM) within the sliding window model, where algorithms utilize limited space to perform large-scale matrix multiplication in a streaming manner. This model has garnered increasing attention in the fields of machine learning and data mining due to its ability to handle time sensitivity and reduce the impact of outdated data. However, despite recent advancements, determining the optimal space bound for this problem remains an open question. In this paper, we introduce the DS-COD algorithm for AMM over sliding windows. This novel and deterministic algorithm achieves optimal performance regarding the space-error tradeoff. We provide theoretical error bounds and the complexity analysis for the proposed algorithm, and establish the corresponding space lower bound for the AMM sliding window problem. Additionally, we present an adaptive version of DS-COD, termed aDS-COD, which improves computational efficiency and demonstrates superior empirical performance. Extensive experiments conducted on both synthetic and real-world datasets validate our theoretical findings and highlight the practical effectiveness of our methods.
\end{abstract}




\maketitle

\section{Introduction}
Large-scale matrix multiplication is a computational bottleneck in many machine learning and data analysis tasks, such as regression~\cite{woodruff2014sketching,naseem2010linear}, clustering~\cite{cohen2015dimensionality,dhillon2001co}, online learning~\cite{TPAMI2022Wan,luo2019robust,duchi2011adaptive,agarwal2019efficient}, and principal component analysis (PCA)~\cite{abdi2010principal,hasan2021review,greenacre2022principal}. For PCA, the product of a high-dimensional data covariance matrix and its eigenvectors typically has a computational complexity of $O(n^3)$, which becomes prohibitive as data size increases. Approximate Matrix Multiplication (AMM)~\cite{drineas2006fast,woodruff2014sketching,ye2016frequent,mroueh2017co} provides an efficient tradeoff between precision and space-time complexity. AMM has proven effective across various tasks~\cite{cohen1999approximating,kyrillidis2014approximate,ye2016frequent,gupta2018oversketch,plancher2019application}. A current trend is to consider AMM in the streaming model, where matrix data is generated in real-time and processed on-the-fly to maintain up-to-date approximations. Among these approaches, the COD algorithm~\cite{mroueh2017co} stands out for its optimal space-error tradeoff, along with superior efficiency and robustness.

In real-world applications, recent data is often prioritized over outdated data. The time sensitivity is common in data analysis tasks, such as analyzing trending topics on social media~\cite{becker2011beyond} over the past week or monitoring network traffic~\cite{joshi2015review} in the last 24 hours. The AMM problem must also account for this time-sensitive nature. For example, in user behavior analysis, researchers are particularly interested in the correlation between user searches and ad clicks in the most recent period.  
The sliding window model~\cite{datar2002maintaining} is well-suited to time-sensitive applications. It processes a data stream by maintaining a window of size $N$, focusing only on the most recent $N$ data items (sequence-based) or the data generated within the most recent $N$ units of time (time-based). The goal of sliding window algorithms is to process only the active data within the window while efficiently discarding outdated information. In most cases, achieving perfectly accurate results requires storing the entire content of the window. This becomes impractical when $N$ or the data dimension is large. Therefore, transitioning from the streaming model to the sliding window model demands novel techniques.

In the AMM over sliding window problem, we have two input matrices $\mX \in \mathbb{R}^{m_x \times n}$ and $\mY \in \mathbb{R}^{m_y \times n}$, and the matrices are received column by column. The active data within the window are represented by $\mX_W \in \mathbb{R}^{m_x \times N}$ and $\mY_W \in \mathbb{R}^{m_y \times N}$. The algorithm processes the incoming columns sequentially while maintaining two low-rank approximations, $\mA \in \mathbb{R}^{m_x \times \ell}$ and $\mB \in \mathbb{R}^{m_y \times \ell}$, where $\ell$ is much smaller than $N$. Its goal is to ensure that the product $\mA\mB^\top$ closely approximates $\mX_W\mY_W^\top$ (i.e., $\mA\mB^\top \approx \mX_W\mY_W^\top$) through the progress, while minimizing both space and time costs. \citet{yao2024approximate} were the pioneers in investigating streaming AMM algorithms over sliding windows. Their proposed EH-COD and DI-COD algorithms are based on the exponential histogram framework~\cite{datar2002maintaining} and the dyadic intervals framework~\cite{arasu2004approximate}. The space complexities of EH-COD and DI-COD are $O\left(\frac{m_x+m_y}{\varepsilon^2}\log{(\varepsilon NR)}\right)$  and $O\left(\frac{m_x+m_y}{\varepsilon}R\log^2{\frac{R}{\varepsilon}}\right)$ respectively, where $R$ represents the upper bound of the squared norm of the data columns. Although their methods are computationally efficient, whether they achieve optimal space complexity remains an open question. \citet{yin2024optimal} proposed an optimal algorithm for matrix sketching over sliding windows, which can be regarded as a symmetric case of AMM in this context; however, developing an optimal algorithm for general AMM over sliding windows continues to be a challenge.

\begin{table}[t]
    \belowrulesep=0pt
    \aboverulesep=0pt
    \centering
    \renewcommand{\arraystretch}{1.5} 
    \caption{Comparison of space complexity of AMM algorithms over sliding windows, with upper bound of relative correlation error $\varepsilon$. Here, $N$ is the window size, $m_x$ and $m_y$ are the dimensions of input matrices $\mX$ and $\mY$, and $R$ is an upper bound on column norms, satisfying $1 \leq \|\vx\|_2\|\vy\|_2 \leq R$.}
    \label{table:comparison}
    \begin{tabular}{|c|c|c|}
        \toprule  
        Methods & Sequence-based & Time-based \\ \hline
        Sampling~\cite{yao2024approximate} & $\frac{m_{x}+m_{y}}{\varepsilon^2} \log (NR)$ & $\frac{m_{x}+m_{y}}{\varepsilon^2} \log (NR)$ \\ 
        EH-COD~\cite{yao2024approximate} & $\frac{m_{x}+m_{y}}{\varepsilon^2} \log (\varepsilon NR)$ & $\frac{m_{x}+m_{y}}{\varepsilon^2} \log (\varepsilon NR)$ \\ 
        DI-COD~\cite{yao2024approximate} & $\frac{m_{x}+m_{y}}{\varepsilon} R \log^2{\frac{R}{\varepsilon}}$ & - \\ 
        \hline
        DS-COD (ours) & $\frac{m_{x}+m_{y}}{\varepsilon} (\log R+1)$ & $\frac{m_{x}+m_{y}}{\varepsilon} \log (\varepsilon NR)$ \\ 
        Lower bound (ours) & $\frac{m_{x}+m_{y}}{\varepsilon} (\log R+1)$ & $\frac{m_{x}+m_{y}}{\varepsilon} \log (\varepsilon NR)$ \\
        \bottomrule
    \end{tabular}
\end{table}

In this paper, we establish a space lower bound for the AMM over the sliding window problem. We prove that any deterministic algorithm for sequence-based sliding windows that achieves an \( \varepsilon \)-approximation error bound requires at least \( \Omega\left(\frac{m_x + m_y}{\varepsilon} (\log{R}+1)\right) \) space. We also propose a new deterministic algorithm, called Dump Snapshot Co-occurring Directions (DS-COD), which utilizes the concept of \( \lambda \)-snapshots \cite{lee2006simpler} and employs a hierarchical structure to handle data from windows with different distributions. Our theoretical analysis shows that in the sequence-based model, the DS-COD algorithm achieves the \( \varepsilon \)-approximation error bound while requiring only \( O\left(\frac{m_x + m_y}{\varepsilon} (\log{R}+1)\right) \) space, demonstrating that it is optimal in terms of space complexity. In the time-based sliding windows, DS-COD requires at most \( O\left(\frac{m_x + m_y}{\varepsilon} \log{(\varepsilon N R)}\right) \) space, matching the space lower bound established for time-based model.
Table~\ref{table:comparison} summarize the comparison of space complexity of existing AMM algorithms over sliding windows.
We also propose an improved algorithm that replaces the multi-level structure with a dynamic adjustment technique, resulting in enhanced performance in experiments and eliminating the need for prior knowledge of $R$. Our main contributions are summarized as follows:
\begin{itemize}
    \item We develop  the DS-COD algorithm to address the AMM over sliding window problem, providing both error bounds and a detailed analysis of space and time complexities.
    \item We establish the lower bounds on the space required by any deterministic algorithm to achieve an $\varepsilon$-approximation error bound in both sequence-based and time-based AMM over sliding window problem. The lower bounds demonstrate that our DS-COD algorithm is optimal in terms of the space-error tradeoff.
    \item Additionally, we propose an enhanced algorithm, aDS-COD, with a dynamic structure that significantly improves the practical computational efficiency, demonstrating strong and stable performance in experiments.
    \item We conduct extensive experiments on both synthetic and real-world datasets, validating our theoretical analysis and demonstrating the effectiveness of our methods.
\end{itemize}

\section{Related Work}
In this section, we present the background and related algorithms for AMM over sliding window.
\paragraph{Streaming AMM}
Over the past few decades, extensive research on AMM has led to the development of various approaches. Randomized methods, such as random sampling~\cite{drineas2006fast} and random projection~\cite{sarlos2006improved, cohen2015optimal}, are simple and effective algorithms that offer better time complexity. Deterministic algorithms, such as FD-AMM~\cite{ye2016frequent} and COD~\cite{mroueh2017co}, base on singular value decomposition and generally achieve smaller approximation errors. In recent years, several variants and extensions of the COD algorithm have been developed. \citet{wan2021approximate} and \citet{luo2021revisiting} proposed optimizations for the sparse data, which improved the costs of time and space. \citet{blalock2021multiplying} introduced learning-augmented methods, providing a better speed-quality tradeoff. \citet{francis2022practical} proposed a variant that is more robust to noise.
\paragraph{Sliding window methods}
The sliding window problem has been a long-standing research focus. One of the most well-known frameworks, proposed by \citet{datar2002maintaining}, uses exponential histograms. By creating a logarithmic hierarchy, it effectively approximates various statistical tasks within the window, such as counting, summing, and computing vector norms. \citet{arasu2004approximate} introduced a tree-based framework with significant practical value. \citet{lee2006simpler} developed a sampling method named $\lambda$-snapshot to approximate frequent items in the sliding window. \citet{braverman2020near} applied random sampling to solve numerical linear algebra problems like spectral and low-rank approximations in the sliding window, achieving near-optimal space efficiency. \citet{wei2016matrix} were the first to study matrix sketching over sliding windows, offering an $\varepsilon$-error approximation using logarithmic space. Recently, \citet{yin2024optimal} proposed a new matrix sketching algorithm based on the $\lambda$-snapshot idea, optimizing matrix sketching over sliding windows.

\paragraph{AMM over sliding windows}
Research on AMM in sliding window models is limited. \citet{yao2024approximate} combined classic sliding window frameworks with AMM, presenting the first comprehensive solution. The EH-COD algorithm, based on the Exponential Histogram idea~\cite{datar2002maintaining}, approximates the matrix product within the window by dividing it into blocks and creating a logarithmic hierarchy with exponentially increasing norms through merge operations. EH-COD works for both sequence-based and time-based windows, requiring $O\left(\frac{m_x+m_y}{\varepsilon^2}\log{(\varepsilon NR)}\right)$ space for sequence-based windows. Its amortized runtime per column is $O\left(\frac{m_x+m_y}{\varepsilon}\log{(\varepsilon NR)}\right)$.

The DI-COD algorithm, using the Dyadic Interval method~\cite{arasu2004approximate}, maintains $L = \log{\frac{R}{\varepsilon}}$ parallel levels to approximate the matrix product with different block granularities. Its space complexity is $O\left(\frac{m_x+m_y}{\varepsilon}R\log^2{\frac{R}{\varepsilon}}\right)$, with an amortized runtime of $O\left(\frac{m_x+m_y}{\varepsilon}\log{\frac{R}{\varepsilon}}\right)$. DI-COD offers better space efficiency than EH-COD when $R$ is small. However, it requires prior knowledge of $R$ and cannot handle variable-size windows, limiting its application to time-based windows. While both EH-COD and DI-COD effectively solve the AMM over sliding window problem, their space efficiency is not optimal.

\section{Preliminary}
In this section, we first present the notation and the problem definition. Then we introduce the COD algorithm and its theoretical guarantees.

\subsection{Notations}
Let $\mI_n$ denote the identity matrix of size $n \times n$, and $\mathbf{0}_{m \times n}$ represent the $m \times n$ matrix filled with zeros. A matrix $\mX$ of size $m \times n$ can be expressed as $\mX = [\vx_1, \vx_2, \dots, \vx_n]$, where each $\vx_i \in \mathbb{R}^m$ is the $i$-th column of $\mX$. The notation $[\mX_1\quad \mX_2]$ represents the concatenation of matrices $\mX_1$ and $\mX_2$ along their column dimensions. For a vector $\vx \in \mathbb{R}^d$, we define its $\ell_2$-norm as $\|\vx\| = \sqrt{\sum_{i=1}^d x_i^2}$. For a matrix $\mX \in \mathbb{R}^{m \times n}$, its spectral norm is defined as $\|\mX\|_2 = \max_{\vu: \|\vu\| = 1} \|\mX \vu\|$, and its Frobenius norm is $\|\mX\|_F = \sqrt{\sum_{i=1}^n \|\vx_i\|^2}$, where $\vx_i$ is the $i$-th column of $\mX$. The condensed singular value decomposition (SVD) of $\mX$, written as SVD$(\mX)$, is given by $\mU \mSigma \mV^\top$, where $\mU \in \mathbb{R}^{m \times r}$ and $\mV \in \mathbb{R}^{n \times r}$ are orthonormal column matrices, and $\mSigma$ is a diagonal matrix containing the nonzero singular values $\sigma_1(\mX) \geq \sigma_2(\mX) \geq \dots \geq \sigma_r(\mX) > 0$. The QR decomposition of $\mX$, denoted as QR$(\mX)$, is given by $\mQ \mR$, where $\mQ \in \mathbb{R}^{m \times n}$ is an orthogonal matrix with orthonormal columns, and $\mR \in \mathbb{R}^{n \times n}$ is an upper triangular matrix. The LDL decomposition is a variant of the Cholesky decomposition that decomposes a positive semidefinite symmetric matrix \( \mX \in \mathbb{R}^{n\times n}\) into \( \mL \mD \mL^\top = \operatorname{LDL}(\mX) \), where \( \mL \) is a unit lower triangular matrix and \( \mD \) is a diagonal matrix. By defining \( \bar{\mL} = \sqrt{\mD} \mL^\top \), we obtain the triangular matrix decomposition \( \mX = \bar{\mL} \bar{\mL}^\top \).

\subsection{Problem Setup}
We first provide the definition of correlation sketch as follows:
\begin{defn}[\cite{mroueh2017co}]
Let $\mX \in \mathbb{R}^{m_x \times n}$, $\mY \in \mathbb{R}^{m_y \times n}$, $\mA \in \mathbb{R}^{m_x \times \ell}$ and $\mB \in \mathbb{R}^{m_y \times \ell}$ where $n\ge\max(m_x,m_y)$ and $\ell \leq \min(m_x,m_y)$. We call the pair $(\mA,\mB)$ is an $\varepsilon$-correlation sketch of $(\mX,\mY)$ if the correlation error satisfies
\[ \text{corr-err}\left(\mX \mY^\top, \mA \mB^\top\right)\triangleq\frac{\norm{\mX\mY^\top-\mA\mB^\top}}{\Norm{\mX}_F \Norm{\mY}_F}\leq \varepsilon. \]
\end{defn}
This paper addresses the problem of approximate matrix multiplication (AMM) in the context of sliding windows. At each time step $t$, the algorithm receives column pairs $(\vx_t, \vy_t)$ from the original matrices $\mX$ and $\mY$. Let $N$ denote the window size. The submatrices within current window are denoted as $\mX_W$ and $\mY_W$. The goal of the algorithm is to maintain a pair of low-rank matrices $(\mA, \mB)$, which is an $\varepsilon$-correlation sketch of the matrices $(\mX_W, \mY_W)$. Similar to \cite{wei2016matrix}, we assume that the squared norms of the data columns are normalized to the range \([1, R]\) for both \( \mX \) and \( \mY \). Therefore, for any column pair \( (\vx, \vy) \), the condition \( 1 \leq \|\vx\| \|\vy\| \leq R \) holds.

\subsection{Co-occurring Directions}
Co-occurring directions (COD)~\cite{mroueh2017co} is a deterministic algorithm for correlation sketching. The core step of COD are summarized in Algorithm \ref{alg:cs}, which we call it the correlation shrinkage (CS) procedure.
\begin{algorithm}[t]
	\renewcommand{\algorithmicrequire}{\textbf{Input:}}
	\renewcommand{\algorithmicensure}{\textbf{Output:}}
	\caption{Correlation Shrinkage (CS)}
	\label{alg:cs}
	\begin{algorithmic}[1]
        \Require
            $\mathbf{A} \in \mathbb{R}^{m_x \times \ell'}, \mathbf{B} \in \mathbb{R}^{m_y \times \ell'}, \text{sketch size }\ell $.
        \State $[\mathbf{Q}_x, \mathbf{R}_x] \leftarrow \text{QR}(\mathbf{A})$,
         $[\mathbf{Q}_y, \mathbf{R}_y] \leftarrow \text{QR}(\mathbf{B})$.
        \State $[\mathbf{U}, \mathbf{\Sigma}, \mathbf{V}] \leftarrow \text{SVD}(\mathbf{R}_x \mathbf{R}_y^\top)$.
        \State $\mC \leftarrow \mQ_x\mU\sqrt{\mathbf{\Sigma}},\mD \leftarrow \mQ_y\mV\sqrt{\mathbf{\Sigma}}$ \Comment{$\mC$ and $\mD$ not computed.}
        \State $\delta \leftarrow \sigma_{\ell} (\mathbf{\Sigma})$,
        $\mathbf{\hat{\Sigma}} \leftarrow \text{max}(\mathbf{\Sigma} - \delta \mathbf{I}_{\ell'}, \mathbf{0})$.
        \State $\mathbf{A} \leftarrow \mathbf{Q}_x \mathbf{U} \sqrt{\mathbf{\hat{\Sigma}}}$,  $\mathbf{B} \leftarrow \mathbf{Q}_y \mathbf{V} \sqrt{\mathbf{\hat{\Sigma}}}$.
        \Ensure 
            $\mathbf{A} \text{ and } \mathbf{B}$.
	\end{algorithmic}  
\end{algorithm}

The COD algorithm initially set $\mA = \mathbf{0}_{m_x \times \ell}$ and $\mB = \mathbf{0}_{m_y \times \ell}$. Then, it processes the i-th column of X and Y as follows
\begin{flalign}
    &\text{Insert $\vx_i$ into a zero valued column of $\mA$} \nonumber\\
    &\text{Insert $\vy_i$ into a zero valued column of $\mB$} \nonumber\\
    &\text{\textbf{if} $\mA$ or $\mB$ has no zero valued columns \textbf{then}} \nonumber\\
    &\text{\quad\quad$[\mA,\mB] = \operatorname{CS}(\mA,\mB, \ell/2)$} \nonumber
\end{flalign}

The COD algorithm runs in $O(n(m_x + m_y)\ell)$ time and requires a space of $O((m_x+m_y)\ell)$. It returns the final sketch $\mA\mB^\top$ with correlation error bounded as:
\[
\norm{\mX\mY^\top-\mA\mB^\top} \leq \frac{2}{\ell}\|\mX\|_F\|\mY\|_F.
\]

For the convenience of expression, we present the following definition:
\begin{defn}
We call matrix pair $(\mC,\mD)$ is an aligned pair of $(\mA,\mB)$ if it satisfies $\mC\mD^\top=\mA\mB^\top$, $\mC=\mQ_x\mU\sqrt{\mathbf{\Sigma}}$ and $\mD=\mQ_y\mV\sqrt{\mathbf{\Sigma}}$, where $\mQ_x$, $\mQ_y$, $\mU$ and $\mV$ are orthonormal matrices and $\mathbf{\Sigma}$ is a diagonal matrix with descending diagonal elements.
\end{defn}
Notice that the line 1-3 of the CS procedure generate an aligned pair $(\mC,\mD)$ for $(\mA,\mB)$. In addition, The output of CS algorithm is a shrinked variant of the aligned pair.

\section{Methods}
\begin{figure}[t]
    \centering
    \includegraphics[width=1\linewidth]{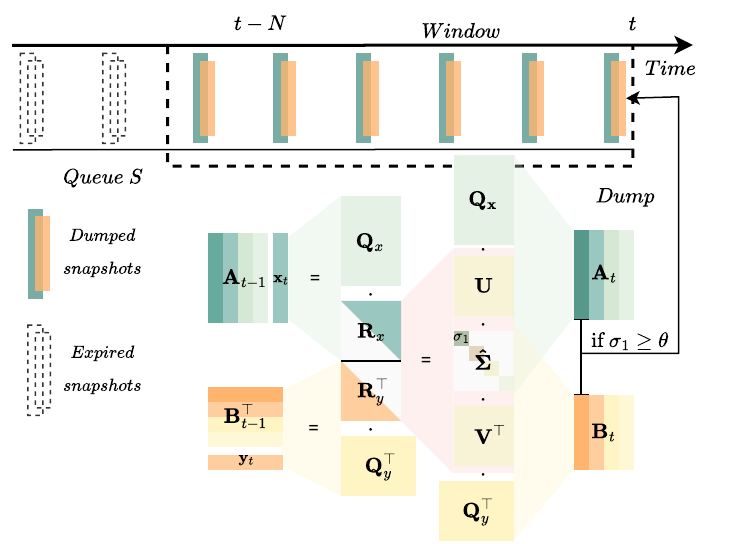}
    \caption{The illustration of Dump Snapshots Co-occurring Directions.}
    \label{fig:idea}
\end{figure}

In this section, we present a novel method that addresses AMM over sliding window, termed Dump Snapshots Co-occurring Directions (DS-COD) which leverage the idea from the $\lambda$-snapshot method~\cite{lee2006simpler} for the frequent item problem in the sliding window model and combine it with the COD algorithm. The main idea of our DS-COD method in figure \ref{fig:idea}. Notice that in the COD algorithm, each CS procedure produces a shrinked aligned pair $\mQ_x \mU \sqrt{\hat{\mathbf{\Sigma}}}$ and $\mQ_y \mV \sqrt{\hat{\mathbf{\Sigma}}}$. Our idea is to examine whether the singular values of $\hat{\mathbf{\Sigma}}$ is large enough. Once a singular value $\sigma_j$ exceeds the threshold $\theta$, we dump the corresponding columns of the aligned pair, i.e. $\sqrt{\sigma_j} \mQ_x \vu_j$ and $\sqrt{\sigma_j} \mQ_y \vv_j$, attaching a timestamp to form a snapshot. 



\subsection{Algorithm Description} \label{sec:alg}

\begin{algorithm}[t]
	\renewcommand{\algorithmicrequire}{\textbf{Input:}}
	\renewcommand{\algorithmicensure}{\textbf{Output:}}
	\caption{DS-COD:Initialize($\ell,\theta$)}
	\label{alg:initialize}
	\begin{algorithmic}[1]
        \Require
            The sketch size $\ell$ and dump threshold $\theta$.
        \State Residual sketches $\mA \leftarrow \mathbf{0}_{m_x\times2\ell},\mB \leftarrow \mathbf{0}_{m_y\times2\ell}$.
        \State Covariance matrices $\mK_A \leftarrow \mathbf{0}_{2\ell\times2\ell},\mK_B \leftarrow \mathbf{0}_{2\ell\times2\ell}$.
        \State Queue of snapshots $S\leftarrow \text{queue}.\operatorname{Initialize}()$.
        \State Variable $\psi \leftarrow 0$.
        
	\end{algorithmic}  
\end{algorithm}

\begin{algorithm}[t]
	\renewcommand{\algorithmicrequire}{\textbf{Input:}}
	\renewcommand{\algorithmicensure}{\textbf{Output:}}
	\caption{DS-COD: Update($\vx_i,\vy_i$)}
	\label{alg:update}
	\begin{algorithmic}[1]
        \Require column pair $\vx_i,\vy_i$, current timestamp $i$
        \State $\psi \leftarrow \psi+\|\vx_i\|\|\vy_i\|$.
        \If{$\operatorname{columns}(\mA)+1\geq 2\ell$}
            \State $\mA,\mB \leftarrow \operatorname{CS}([\mA \quad \vx_i],[\mB \quad \vy_i],\ell)$.
             \While{$\|\va_1\|\|\vb_1\|\geq \theta$}
                \State $S$ append snapshot$(v=(\va_1,\vb_1),t=i)$.
                \State Remove $\va_1,\vb_1$ from $\mA,\mB$ respectively.
            \EndWhile
            \State $\mK_A=\mA^\top\mA,\mK_B=\mB^\top\mB$.
            \State $\psi \leftarrow \|\va_1\|\|\vb_1\|$.
        \Else
            \State $\mK_A\leftarrow\begin{bmatrix}\mK_A&\mA^\top\vx_i\\\vx_i^\top\mA&\vx_i^\top\vx_i\end{bmatrix},\mK_B\leftarrow\begin{bmatrix}\mK_B&\mB^\top\vy_i\\\vy_i^\top\mB&\vy_i^\top\vy_i\end{bmatrix}$.
            \State $\mA\leftarrow [\mA;\vx_i],\mB\leftarrow [\mB;\vy_i]$.
            \If{$\psi \geq \theta$}
                 \State $[\mL_A,\mD_A]\leftarrow \operatorname{LDL}(\mK_A), [\mL_B,\mD_B]\leftarrow \operatorname{LDL}(\mK_B)$.
                 \State $\mR_x \leftarrow \sqrt{\mD_A}\mL_A^\top,\mR_y \leftarrow \sqrt{\mD_B}\mL_B^\top$.
                \State $[\mU,\mathbf{\Sigma},\mV] \leftarrow \operatorname{SVD}(\mR_x\mR_y^\top)$.
                \For{$j = 1,2,\dots,\ell$}
                    \If{$\sigma_j \geq \theta$}
                        \State $\va_j\leftarrow \frac{1}{\sqrt{\sigma_j}}\mA\mR_y^{T}\vv_j,\vb_j\leftarrow \frac{1}{\sqrt{\sigma_j}}\mB\mR_x^{T}\vu_j$.
                        \State $S$ append snapshot$(s=(\va_j,\vb_j),t=i)$.
                        \State $\mP_A\leftarrow \frac{1}{\sigma_j}\mR_y^{T}\vv_j\vu_j^\top\mR_x,\mP_B\leftarrow \frac{1}{\sigma_j}\mR_x^{T}\vu_j\vv_j^\top\mR_y$.
                        \State $\mA \leftarrow \mA - \mA\mP_A,\mB \leftarrow \mB - \mB\mP_B$.
                        \State $\mK_A \leftarrow (\mI-\mP_A^\top)\mK_A(\mI-\mP_A)$.
                        \State $\mK_B \leftarrow (\mI-\mP_B^\top)\mK_B(\mI-\mP_B)$.
                    \Else
                        \State BREAK.
                    \EndIf
                \EndFor
            \EndIf
            \State $\psi \leftarrow \sigma_j$.
            
        \EndIf
        
        \Ensure $\mA,\mB$ and $S$.
	\end{algorithmic}  
\end{algorithm}

Algorithm \ref{alg:initialize} illustrates the initialization of a DS-COD sketch. The DS-COD sketch maintains two residual matrices, \( \mA \in \mathbb{R}^{m_x \times 2\ell} \) and \( \mB \in \mathbb{R}^{m_y \times 2\ell} \), to receive newly arriving columns. \( \mK_A \in \mathbb{R}^{2\ell \times 2\ell} \) and \( \mK_B \in \mathbb{R}^{2\ell \times 2\ell} \) represent the covariance matrices of \( \mA \) and \( \mB \), respectively. The snapshot queue \( \mathcal{S} \) is initialized as empty, and the variable \( \psi \) is initialized to 0 to track the maximum singular value of \( \mA \mB^\top \).
Algorithm \ref{alg:update} outlines the pseudocode for updating a DS-COD sketch. The update process involves two cases: perform a CS operation once the buffer is full, or do a quick-check when the buffer has fewer than $2\ell$ columns. These correspond to lines 2-8 (case 1) and lines 10-26 (case 2), respectively. It is important to emphasize that timely snapshot dumping is essential. One direct approach is to perform a CS operation after each update. However, since the time to run a CS procedure is $O((m_x + m_y)\ell^2)$, frequent CS operations are prohibitive. Therefore, in case 2 we introduces a quick-check approach.
The DS-COD algorithm maintains the covariance matrices \( \mK_A = \mA^\top \mA \) and \( \mK_B = \mB^\top \mB \), updating them in a rank-1 manner (line 10). Since $\mA^\top \mA = \mR_x^\top \mQ_x^\top \mQ_x \mR_x = \mR_x^\top \mR_x$, the LDL decomposition of $\mK_A$ can quickly yields $\mR_x=\sqrt{\mD}\mL^\top$. Also, $\mR_y$ can be obtained by the same process (line 14). By this way we can avoid time-consuming QR decomposition.

 Lines 16-23 inspect the singular values one by one, compute the snapshots to be dumped, and eliminate their influence from the buffers $\mA$, $\mB$, and the corresponding covariance matrices $\mK_A$, $\mK_B$. Recall that a snapshot consists of columns from the aligned pair. However, since the quick-check does not compute the complete aligned pair, we need to calculate the required columns individually. For instance, if $\sigma_1$ exceeds the threshold $\theta$, then the snapshot takes the first column of the aligned pair of $\mA\mB^\top$, which should be $\sqrt{\sigma_1} \mQ_x \vu_1$ and $\sqrt{\sigma_1} \ \mQ_y \vv_1$. 
In the absence of $\mQ_x$ and $\mQ_y$, we use an alternative expression for the aligned pair (instead of $\mathbf{A} \mR_x^{-1} \mathbf{U} \sqrt{\mathbf{\Sigma}}$ to save time on matrix inversion): 
\begin{align} \label{eq:rm}
\mQ_x\mU\sqrt{\mathbf{\Sigma}}=\mathbf{A}\mathbf{R}_y^\top\mathbf{V}\sqrt{\mathbf{\Sigma}}^{-1},\quad \mQ_y\mV\sqrt{\mathbf{\Sigma}}=\mathbf{B}\mathbf{R}_x^\top\mathbf{U}\sqrt{\mathbf{\Sigma}}^{-1}.
\end{align}
This demonstrates the form of the snapshot as described in line~18.
If we had the aligned pair, we could simply remove the first column to complete the dump. However, since we only have the original matrices $\mA$ and $\mB$, removing the snapshot's impact from them presents significant challenges. To reconsider the problem: we have buffers $\mA$ and $\mB$, and suppose $\mC$ and $\mD$ are the aligned pair of $\mA \mB^\top$, such that $\mA \mB^\top = \mQ_x \mU \mathbf{\Sigma} \mV^\top \mQ_y^\top = \mC \mD^\top$. Let $\overline{\mC}$ and $\overline{\mD}$ represent the results of removing the first column from $\mC$ and $\mD$, respectively. The remain question is how to update $\mA$ and $\mB$ so that $\overline{\mA}_{} \overline{\mB}_{}^\top = \overline{\mC}_{} \overline{\mD}_{}^\top$. The approach is provided in lines 20-21, and we explain it in the proof of Lemma \ref{lem:remove} (in appendix \ref{apdx:remove}).

\begin{lem} \label{lem:remove}
    Let $[\mQ_x,\mR_x]=\operatorname{QR}(\mA)$, $[\mQ_y,\mR_y]=\operatorname{QR}(\mB)$
    , \( [\mU, \mathbf{\Sigma}, \mV] = \operatorname{SVD}(\mR_x \mR_y^\top) \), \( \mC = \mQ_x \mU \sqrt{\mathbf{\Sigma}} \) and \( \mD = \mQ_y \mV \sqrt{\mathbf{\Sigma}} \). Then $(\mC,\mD)$ is an aligned pair of $(\mA,\mB)$. If we construct $\overline{\mC}$ and $\overline{\mD}$ by removing the \( j \)-th column of $\mC$ and $\mD$, respectively, then $(\overline{\mC},\overline{\mD})$ is an aligned pair of of $(\overline{\mA},\overline{\mB})$, where
    
    \[
    \overline{\mA} = \mA - \mQ_x \vu_j \vu_j^\top \mR_x, \quad \overline{\mB} = \mB - \mQ_y \vv_j \vv_j^\top \mR_y.
    \]
\end{lem}
According to \eqref{eq:rm}, we obtain $\mQ_x = \mathbf{A} \mathbf{R}_y^\top \mathbf{V} \mathbf{\Sigma}^{-1} \mathbf{U}^\top$ and $\mQ_y = \mathbf{B} \mathbf{R}_x^\top \mathbf{U} \mathbf{\Sigma}^{-1} \mathbf{V}^\top$. Then, we can express the update as follows:
\[
\mA - \mQ_x \vu_j \vu_j^\top \mR_x = \mA - \mathbf{A} \mathbf{R}_y^\top \mathbf{V} \mathbf{\Sigma}^{-1} \mathbf{U}^\top \vu_j \vu_j^\top \mR_x = \mA - \frac{1}{\sigma_j} \mathbf{A} \mathbf{R}_y^\top \vv_j \vu_j^\top \mR_x.
\]
Similarly, the update for $\mB$ is $\mB - \frac{1}{\sigma_j}\mB \mathbf{R}_x^\top \vu_j \vv_j^\top \mR_y$. In line 23-24, we update $\mK_A$ to $\overline{\mA}^\top \overline{\mA}$ and $\mK_B$ to $\overline{\mB}^\top \overline{\mB}$.

Note that the quick-check is not performed after every update. The variable $\psi$ tracks the residual maximum singular value after snapshots dumped (line 26). During updates, the norm product $\|\vx\| \|\vy\|$ is accumulated into $\psi$ (line 1). Suppose the residual maximum singular value is $\sigma$. After receiving columns $\vx$ and $\vy$, let the aligned pair be $\mC$ and $\mD$, satisfying $\mC \mD^\top = \mA\mB^\top + \vx\vy^\top$. By definition, we have:
\begin{align*}
\|\vc_1\|\|\vd_1\| = \sigma_1(\mA\mB^\top + \vx\vy^\top) \leq \sigma_1(\mA\mB^\top) + \sigma_1(\vx\vy^\top) 
=\sigma + \|\vx\|\|\vy\|.
\end{align*}

Therefore, A snapshot dump only occurs when $\psi$ exceeds the threshold $\theta$.

\subsection{Hierarchical DS-COD}
We now introduce the complete DS-COD algorithm framework (in Algorithm \ref{alg:hdscod}) for sequence-based sliding window  and explain how to use appropriate thresholds $\theta$ to generate enough snapshots to meet the required accuracy, while keeping space usage limited.
\begin{algorithm}[t]
	\renewcommand{\algorithmicrequire}{\textbf{Input:}}
	\renewcommand{\algorithmicensure}{\textbf{Output:}}
	\caption{Hierarchical DS-COD (hDS-COD)}
	\label{alg:hdscod}
	\begin{algorithmic}[1]
        \Require
            $\mX \in \mathbb{R}^{m_x \times n},\mY \in \mathbb{R}^{m_y \times n}$,  window size $N$, sketch size $\ell$.
        \renewcommand{\algorithmicrequire}{\textbf{Initialize:}}
        \Require
            Initialize $L+1$ DS-COD sketches in $\mathcal{L}[0]$ to $\mathcal{L}[L]$, with exponentially increasing threshold : $\varepsilon N, 2\varepsilon N,  \dots, 2^L\varepsilon N$. Simultaneously, configure a set of auxiliary sketches \( \mathcal{L}_{aux}[0] \) to \( \mathcal{L}_{aux}[L] \) with the same settings.
        \For{$ i =1,2,\dots,n$}
            \For{$j =0,1,\dots,L$}
                \While{$\mathcal{L}[j].S[0].t + N \leq i$ \textbf{or} $\operatorname{len}(\mathcal{L}[j].S) > \frac{1}{\varepsilon}$}
                    \State $\mathcal{L}[j].S.\operatorname{Dequeue}()$.
                \EndWhile
                \State $\mathcal{L}[j].\operatorname{Update}(\vx_i,\vy_i)$.
                \State $\mathcal{L}_{aux}[j].\operatorname{Update}(\vx_i,\vy_i)$.
                \If{$i \operatorname{mod} N == 1$}
                    \State $\mathcal{L}[j] \leftarrow \mathcal{L}_{aux}[j]$.
                    \State $\mathcal{L}_{aux}[j].\operatorname{Initialize}()$.
                \EndIf
            \EndFor
        \EndFor
	\end{algorithmic}  
\end{algorithm}

Since the squared norms of the columns may vary across different windows but always stay within the range $[1, R]$, we use hierarchical thresholds to capture this variation. 
The algorithm constructs a structure with \( L = \lceil \log_2{R} \rceil \) levels, where the thresholds increase exponentially. Specifically, the threshold at the \( j \)-th level is \( 2^j \varepsilon N \), meaning that lower levels have a higher frequency of snapshot dumping. To prevent lower levels from generating excessive snapshots, potentially up to \( N \), we cap the total number of snapshots in a DS-COD sketch at \( 1/\varepsilon \).

The algorithm begins by discarding expired or excessive snapshots from the head of the queue \( S \) (lines 3-4), and then proceeds with the updates. We store the primary components of the aligned pair with timestamps as snapshots, while the components in the residual matrices \( \mA \) and \( \mB \) also require expiration handling to maintain the algorithm's error bounds. These residual matrices are managed using a coarse expiration strategy, referred to as $N$-restart. Each level also includes an auxiliary sketch \( \mathcal{L}_{aux} \), which is updated alongside the main sketch. After every \( N \) update steps, the algorithm swaps the contents of \( \mathcal{L} \) and \( \mathcal{L}_{aux} \), and then reinitializes the auxiliary sketch (lines 7-9).
\paragraph{Query Algorithm} The query algorithm of Hierarchical DS-COD is presented in Algorithm \ref{alg:query}. 
Since lower layers may not encompass the entire window, the query process starts from the top and searches for a layer with at least $\Omega(\ell)$ non-expiring snapshots. The snapshots are then merged with the residual matrices, and the final aligned pair are obtained using the CS operation.

\begin{algorithm}[t]
	\renewcommand{\algorithmicrequire}{\textbf{Input:}}
	\renewcommand{\algorithmicensure}{\textbf{Output:}}
	\caption{Hierarchical DS-COD: Query($\mathcal{L}$)}
	\label{alg:query}
	\begin{algorithmic}[1]
        \Require
            sketch sequence $\mathcal{L}$.
        \State Find the $\max_j{\operatorname{len}(\mathcal{L}[j].S]) \geq \Omega(\ell)}$.
        \State Stack snapshots in $\mathcal{L}[j].S$ to form $\mA'$ and $\mB'$.
        \State $[\mA,\mB] \leftarrow \operatorname{CS}([\mathcal{L}[j].\mA \quad \mA'],[\mathcal{L}[j].\mB \quad \mB'],\ell)$.
        \Ensure
            $\mA$ and $\mB$.
\end{algorithmic}  
\end{algorithm}

We present the following main results on the correlation error of hDS-COD in AMM over sliding window. The proof of error bound and complexity is given in appendix \ref{apdx:hds} and \ref{apdx:comlexity}, respectively.
\begin{thm} \label{thm:hds} 
    By setting \( \ell = \frac{1}{\varepsilon} \), the hDS-COD algorithm can generate sketches \( \mA \) and \( \mB \) of size \( O(\ell) \), with the correlation error bounded by
    \[
        \text{corr-err }\left(\mX_W\mY_W^\top,\mA\mB^\top\right) \leq 8\varepsilon.
    \]
    The hDS-COD algorithm requires  \( O\left( \frac{m_x + m_y}{\varepsilon} (\log{R}+1) \right) \) space cost and the amortized time cost per column is \( O\left( \frac{m_x + m_y}{\varepsilon} (\log{R}+1) \right) \).
\end{thm}

Compared to existing algorithms, hDS-COD provides a substantial improvement in space complexity. In contrast to EH-COD~\cite{yao2024approximate} which requires  $\frac{m_{x}+m_{y}}{\varepsilon^2} \log (\varepsilon NR)$ space cost, our hDS-COD algorithm reduces the the dependence on $\varepsilon$ from $\frac{1}{\varepsilon^2}$ to $\frac{1}{\varepsilon}$. In contrast to DI-COD~\cite{yao2024approximate} which requires $\frac{m_{x}+m_{y}}{\varepsilon} R \log^2{\frac{R}{\varepsilon}}$ space cost, our hDS-COD algorithm significantly reduce the dependence on $R$. On the other hand, hDS-COD also offers advantages in time complexity. Notably, the time complexities per column of EH-COD and DI-COD are  \( O(\frac{m_x + m_y}{\varepsilon} \log{(\varepsilon NR)}) \) and $O\left( \frac{m_x + m_y}{\varepsilon} \log \frac{R}{\varepsilon} \right)$, respectively. In comparison, hDS-COD maintains a competitive time complexity.

\subsection{Adaptive DS-COD}
The multilayer structure of hDS-COD effectively captures varying data patterns by selecting the appropriate level for each window, providing approximations with a bounded error. However, it requires prior knowledge of the maximum column squared norm $R$, and maintains $\log R$ levels of snapshot queues and residual sketches, with the update cost also increasing with $\log R$. In many cases, the dataset may have a large $R$ but little fluctuation within the window, leading to wasted space and unnecessary updates, especially at extreme levels. To address this issue, we propose a heuristic variant of hDS-COD, called adaptive DS-COD (aDS-COD), in Algorithm \ref{alg:adscod}.
\begin{algorithm}[t]
	\renewcommand{\algorithmicrequire}{\textbf{Input:}}
	\renewcommand{\algorithmicensure}{\textbf{Output:}}
	\caption{Adaptive DS-COD (aDS-COD)}
	\label{alg:adscod}
	\begin{algorithmic}[1]
        \Require
            $\mX \in \mathbb{R}^{m_x \times n},\mY \in \mathbb{R}^{m_y \times n}$,  window size $N$, sketch size $\ell$.
        \renewcommand{\algorithmicrequire}{\textbf{Initialize:}}
        \Require
            Initialize a DS-COD sketch $DS$ and an auxiliary sketch $DS_{aux}$ with initial threshold $\varepsilon N$ and threshold level $L=1$.
        \For{$ i =1,2,\dots,n$}
            \While{$ DS.S[0].t + N \leq i$}
                \State $DS.S.\operatorname{Dequeue}()$.
            \EndWhile
            \If{$i \operatorname{mod} N == 1$}
                \State $DS \leftarrow DS_{aux}$.
                \State $DS_{aux}.\operatorname{Initialize}()$.
            \EndIf
            \State $DS.\operatorname{Update}(\vx_i,\vy_i)$.
            \State $DS_{aux}.\operatorname{Update}(\vx_i,\vy_i)$.
    
            \If{$\operatorname{len}(DS.S) \geq \frac{ L}{\varepsilon}$}\Comment{adjust threshold}
                \State $L \leftarrow L + 1$.
                \State $DS.\theta \leftarrow 2\cdot DS.\theta $.
            \ElsIf{$\operatorname{len}(DS.S) \leq \frac{L-1}{\varepsilon}$}
                \State $L \leftarrow L - 1$.
                \State $DS.\theta \leftarrow DS.\theta / 2 $.
            \EndIf
        \EndFor
	\end{algorithmic}  
\end{algorithm}

The algorithm maintains a single pair of DS-COD sketches: a main sketch $DS$ and an auxiliary sketch $DS_{aux}$. Initially, the algorithm handles expiration without restricting the number of snapshots ( the snapshot count is inherently bounded by $O(\ell \log R)$). The update process follows the same steps as in hDS-COD. Since there is only one level, the initial threshold $\theta$ is set to $\varepsilon N$, and the variable L, representing the current threshold level, is initialized to 1. After each update, the algorithm checks the current number of snapshots.If it exceeds $L/\varepsilon$, this indicates that the threshold is too small, and to prevent an excessive number of snapshots in the future, the threshold is doubled and the threshold level is increased accordingly (lines 9-10). In contrast, if the current number of snapshots is below the lower bound $(L-1)/\varepsilon$, it suggests that the squared norms of the data in the current window are decreasing and the threshold level needs to be lowered. Similarly, \( DS_{aux} \) also requires threshold adjustment; for clarity, this step is omitted from the pseudocode.

For typical datasets, the approximation error of aDS-COD is comparable to that of hDS-COD. Additionally, since aDS-COD employs a single-layer structure, it requires less space than hDS-COD empirically. The space upper bound remains unchanged: in the worst-case scenario, the threshold level increases from $1$ to $\log R$, with each level accumulating $\ell$ snapshots. Therefore, the total number of snapshots is bounded by $O((m_x+m_y)\ell \log R)$. The time complexity for handling updates is $O((m_x+m_y)\ell)$.

\subsection{Time-based model}
Many real-world applications based on sliding windows focus on actual time windows, such as data from the past day or the past hour, which are known as time-based sliding windows. The key difference between time-based models and sequence-based models lies in the variability of data arrival within the same time span, which can sometimes fluctuate dramatically. We assume that at most one data point arrives at any given time point. When no data arrives at a particular time point, the algorithm equivalently processes a pair of zero vectors. Consequently, the assumption  on  column pairs $(\vx,\vy)$ of the input matrices changes to $\|\vx\|\|\vy\| \in [1, R] \cup \{0\}$.

We demonstrate how to adapt hDS-COD to the time-based model. Due to the presence of zero columns, the norm product within a window $\|\mathbf{X}_W\|_F \|\mathbf{Y}_W\|_F$ ranges within $[0, NR]$. Accordingly, the thresholds for generating snapshots need to be adjusted to $\theta_i = 2^i$. For hDS-COD, it is necessary to initialize $\log{(\varepsilon NR)}$ levels with exponentially increasing thresholds from $1$ to $\varepsilon NR$. In contrast, aDS-COD only requires setting the initial threshold to $1$, but its space upper bound remains consistent with that of hDS-COD, at $O((m_x + m_y)\ell \log{(\varepsilon NR)})$.

\begin{corollary} \label{corollary:ds-time} 
    If we set $\ell=\frac{1}{\varepsilon}$, the time-based hDS-COD can generate sketches $\mA$ and $\mB$ of size $O(\ell)$, with the correlation error upper bounded by 
    \[
        \text{corr-err }\left(\mX_W\mY_W^\top,\mA\mB^\top\right) \leq 8\varepsilon.
    \]
    The algorithm uses $O\left(\frac{m_x + m_y}{\varepsilon} \log{(\varepsilon NR)}\right)$ space and requires \\$O\left(\frac{m_x + m_y}{\varepsilon} \log{(\varepsilon NR)}\right)$ update time.
\end{corollary}

\section{Lower Bound}
In this section, we establish the space lower bounds for any deterministic algorithm designed for the problem of AMM over sliding windows. The proposed lower bounds match the space complexities of our DS-COD algorithm and demonstrate that the optimality of the DS-COD algorithm with regard to the memory usage. We begin with the following lemma.
\begin{lem} \label{lem:matrix_set}
    Suppose $m_x\leq m_y$. For any $\ell<m_x$, there exists a set of matrix pairs $\mathcal{Z}_{\ell}$ = $\left\{ ( \mathbf{X}^{(1)}, \mathbf{Y}^{(1)} ), \cdots , ( \mathbf{X}^{(M)}, \mathbf{Y}^{(M)} ) \right\}$ with $M=2^{\Omega((m_x+m_y)\ell)}$,  $\mathbf{X}^{(i)}\in\mathbb{R}^{m_x \times \ell}$ and $\mathbf{Y}^{(i)}\in\mathbb{R}^{m_y \times \ell}$ that satisfies $\mathbf{X}^{(i)\top} \mathbf{X}^{(i)} = \mathbf{Y}^{(i)\top} \mathbf{Y}^{(i)} = \mathbf{I}_{\ell}$ for $i=1, \ldots, M$ and 
    \[
    \left\|\mathbf{X}^{(i)} \mathbf{Y}^{(i)\top} - \mathbf{X}^{(j)} \mathbf{Y}^{(j)\top}\right\|_2 > \frac{1}{2}
    \]
    for any $i \neq j$.
\end{lem}
\begin{proof}
This lemma can be directly obtained from Lemma 6 of \cite{luo2021revisiting} with $\delta=1/8$.
\end{proof}

Then we present the lower bound on the space complexity for approximate matrix multiplication over sequence-based sliding windows. The proof is provided in Appendix \ref{apdx:lowerbound}.
\begin{thm}\label{thm:lowerbound}
    Consider any deterministic algorithm that outputs an $(\varepsilon/9)$-correlation sketch  $(\mathbf{A}_W,\mathbf{B}_W)$ for the matrices $(\mathbf{X}_W,\mathbf{Y}_W)$ in the sequence-based sliding window, 
    where $\mathbf{X}_W \in \mathbb{R}^{(m_x + 1) \times N}$, $\mathbf{Y}_W \in \mathbb{R}^{(m_y + 1) \times N}$ and $N \geq \frac{1}{2\varepsilon} \log_2{\frac{R}{\varepsilon}}$. Assuming that all column pairs $(\vx_i,\vy_i)$ in $\mathbf{X}_W$ and $\mathbf{Y}_W$ satisfies $1 \leq \|\mathbf{x}_i\| \|\mathbf{y}_i\| \leq R + 1$, then the algorithm requires at least $\Omega\left(\frac{m_x + m_y}{\varepsilon} (\log{R}+1)\right)$ bits of space.
\end{thm}

We also establish a space lower bound for the AMM in the time-based sliding window model, with the proof in Appendix \ref{apdx:lowerbound_time}.
\begin{thm}\label{thm:lowerbound_time}
    Consider any deterministic algorithm that outputs an $(\varepsilon/3)$-correlation sketch  $(\mathbf{A}_W,\mathbf{B}_W)$ for the matrices $(\mathbf{X}_W,\mathbf{Y}_W)$ in the time-based sliding window, where $\mathbf{X}_W \in \mathbb{R}^{m_x \times N}$, $\mathbf{Y}_W \in \mathbb{R}^{m_y \times N}$ and $N \geq \frac{1}{2\varepsilon} \log_2{\frac{R}{2}}$. Assuming that all column pairs $(\vx_i,\vy_i)$ in $\mathbf{X}_W$ and $\mathbf{Y}_W$ satisfies $\|\mathbf{x}_i\| \|\mathbf{y}_i\| \in 0\cup[1, R]$, then the algorithm requires at least $\Omega\left(\frac{m_x + m_y}{\varepsilon} \log(\varepsilon NR)\right)$ bits of space.
\end{thm}

\section{Experiments}
In this section, we empirically compare the proposed algorithm on both sequence windows and time windows with existing methods.
\paragraph{Datasets} For the sequence-based model, we used two synthetic datasets and two cross-language datasets. The statistics of the datasets are provided in Table \ref{table:statistics}:

\begin{table}[t]
    \centering
    \caption{The statistics of the datasets. The datasets satisfy $1 \leq \|\vx\|\|\vy\| \leq R $.}
    \label{table:statistics}
    \begin{tabular}{|c|c|c|c|c|c|}
    \hline
        Dataset & $n$ & $m_x$ & $m_y$ & $N$ & $R$ \\ \hline
        SYNTHETIC(1) & 100,000 & 1,000 & 2,000 & 50,000 & 65 \\ \hline
        SYNTHETIC(2) & 100,000 & 1,000 & 2,000 & 50,000 & 724 \\ \hline
        APR & 23,235 & 28,017 & 42,833 & 10,000 & 773 \\ \hline
        PAN11 & 88,977 & 5,121 & 9,959 & 10,000 & 5,548 \\ \hline
        EURO & 475,834 & 7,247 & 8,768 & 100,000 & 107,840 \\ \hline
    \end{tabular}
\end{table}

\begin{itemize}
    \item Synthetic: The elements of the two synthetic datasets are initially uniformly sampled from the range (0,1), then multiplied by a coefficient to adjust the maximum column squared norm $R$. The X matrix has 1,000 rows, and the Y matrix has 2,000 rows, each with 100,000 columns. The window size is set to 50,000.
    \item APR: The Amazon Product Reviews (APR) dataset is a publicly available collection containing product reviews and related information from the Amazon website. This dataset consists of millions of sentences in both English and French. We structured it into a review matrix where the X matrix has 28,017 rows, and the Y matrix has 42,833 rows, with both matrices sharing 23,235 columns. The window size is 10,000.
    \item PAN11: PANPC-11 (PAN11) is a dataset designed for text analysis, particularly for tasks such as plagiarism detection, author identification, and near-duplicate detection. The dataset includes texts in English and French. The X and Y matrices contain 5,121 and 9,959 rows, respectively, with both matrices having 88,977 columns. The window size is 10,000.
\end{itemize}
We evaluate the time-based model on another real-world dataset:
\begin{itemize}
    \item EURO: The Europarl (EURO) dataset is a widely used multilingual parallel corpus, comprising the proceedings of the European Parliament. We selected a subset of its English and French text portions. The X and Y matrices contain 7,247 and 8,768 rows, respectively, and both matrices share 475,834 columns. Timestamps are generated using the $Poisson$ $Arrival$ $Process$ with a rate parameter of $\lambda=2$. The window size is set to 100,000, with approximately 30,000 columns of data on average in each window.
\end{itemize}

\paragraph{Setup} For the sequence-based model, we compare the proposed hDS-COD and  aDS-COD with EH-COD~\cite{yao2024approximate} and DI-COD~\cite{yao2024approximate}. We do not consider the Sampling algorithm as a baseline, as its performance is inferior to that of EH-COD and DI-CID, as demonstrated in \cite{yao2024approximate}. 
For the time-based model, we compare the proposed hDS-COD and  aDS-COD with EH-COD and the Sampling algorithm since DI-COD cannot be applied to time-based sliding window model. To achieve the same error bound, the maximum number of levels for hDS-COD is set to $L = \log{(\varepsilon NR)}$, and the initial threshold for aDS-COD is set to $1$.

Our experiments aim to illustrate the trade-offs between space and approximation errors. The x-axis represents two metrics for space: final sketch size and total space cost. The final sketch size refers to the number of columns in the result sketches $\mA$ and $\mB$ generated by the algorithm, representing a compression ratio. The total space cost refers to the maximum space required during the algorithm's execution, measured by the number of columns.We evaluate the approximation performance of all algorithms based on correlation errors $\operatorname{corr-err}(\mathbf{X}_W \mathbf{Y}_W^\top, \mathbf{A} \mathbf{B}^\top)$, which is reflected on the y-axis. Every 1,000 iterations, all algorithms query the window and record the average and maximum errors across all sampled windows.

The experiments for all algorithms were conducted using MATLAB (R2023a), with all algorithms running on a Windows server equipped with 32GB of memory and a single processor of Intel i9-13900K.

\begin{figure*}[t]


    \subfigure[SYNTHETIC(1)]{\includegraphics[width=0.25\textwidth]{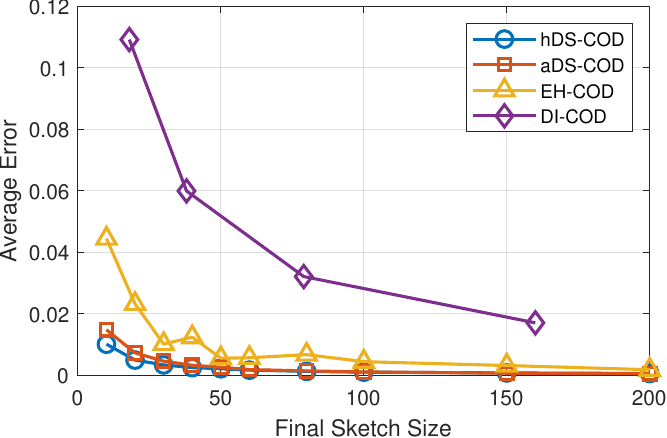}}\hfill
    \subfigure[SYNTHETIC(2)]{\includegraphics[width=0.25\textwidth]{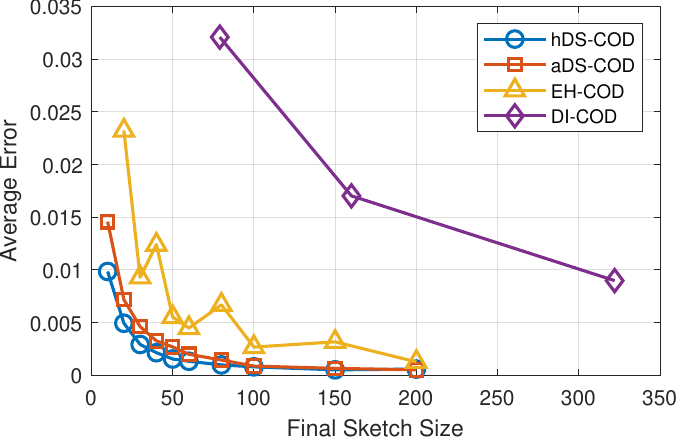}}\hfill
    \subfigure[APR]{\includegraphics[width=0.25\textwidth]{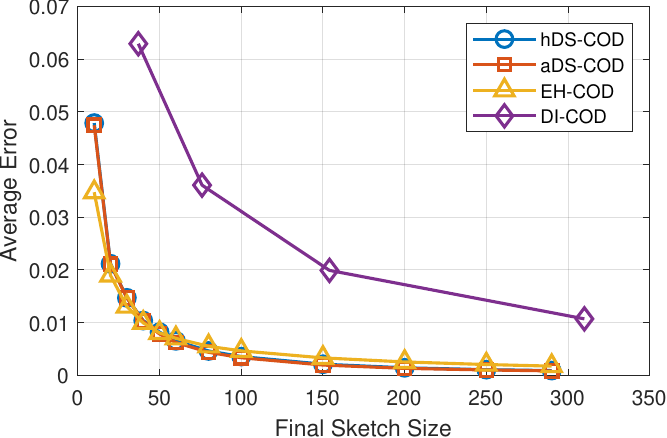}}\hfill
    \subfigure[PAN11]{\includegraphics[width=0.25\textwidth]{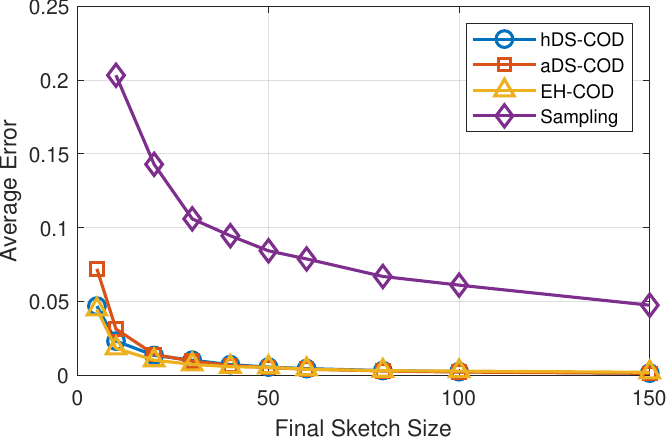}}\hfill

\vspace{-0.2cm}

   \subfigure[SYNTHETIC(1)]{\includegraphics[width=0.25\textwidth]{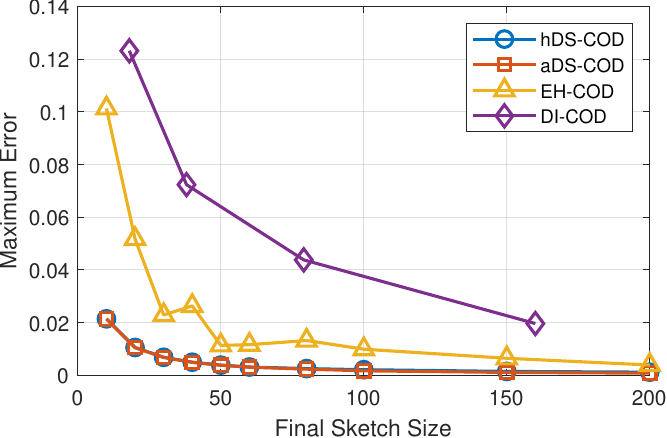}}\hfill
    \subfigure[SYNTHETIC(2)]{\includegraphics[width=0.25\textwidth]{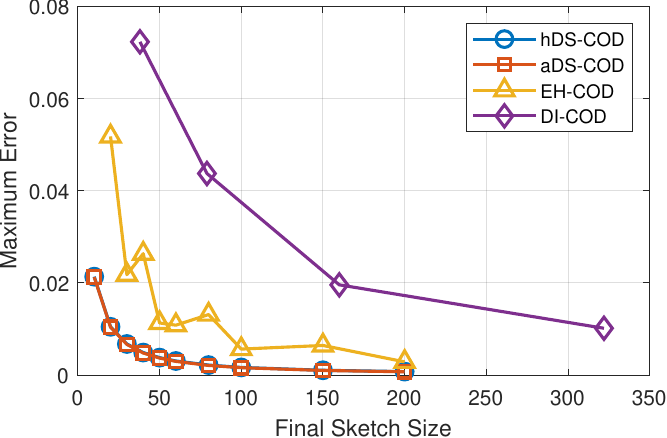}}\hfill
    \subfigure[APR]{\includegraphics[width=0.25\textwidth]{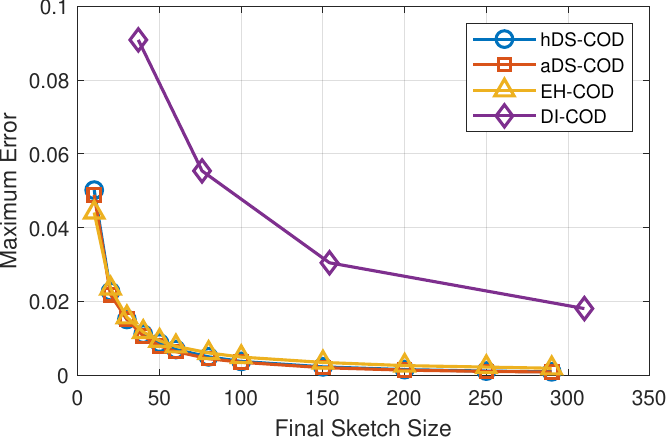}}\hfill
    \subfigure[PAN11]{\includegraphics[width=0.25\textwidth]{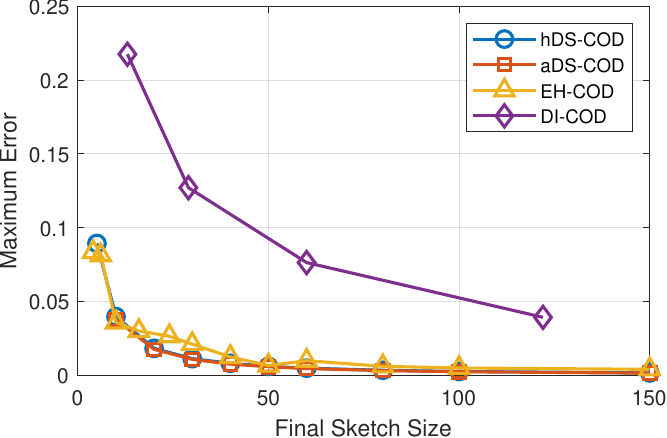}}\hfill
    
\vspace{-0.3cm}
    \caption{The plot of correlation error against final sketch size.}
    \label{fig:error vs l}
    
\vspace{0.5cm}

    \subfigure[SYNTHETIC(1)]{\includegraphics[width=0.25\textwidth]{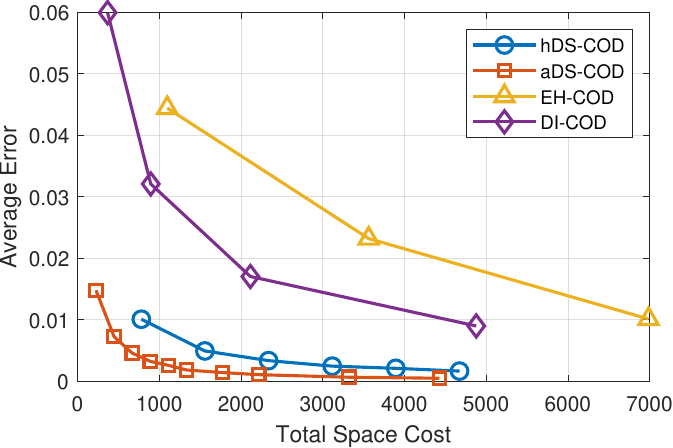}}\hfill
    \subfigure[SYNTHETIC(2)]{\includegraphics[width=0.25\textwidth]{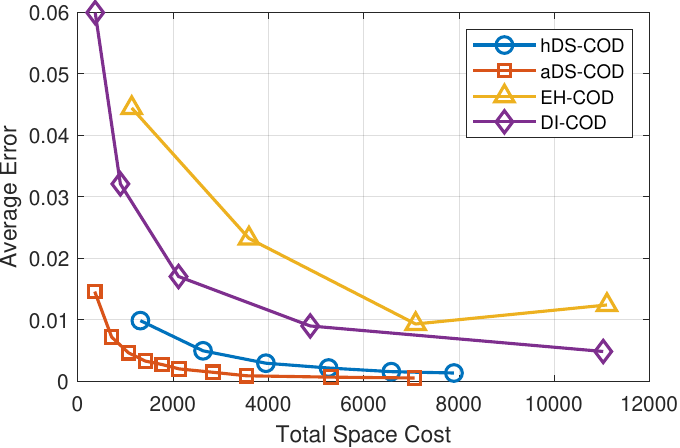}}\hfill
    \subfigure[APR]{\includegraphics[width=0.25\textwidth]{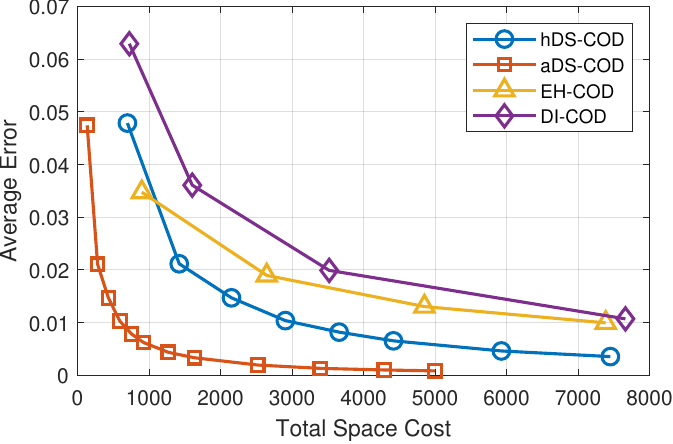}}\hfill
    \subfigure[PAN11]{\includegraphics[width=0.25\textwidth]{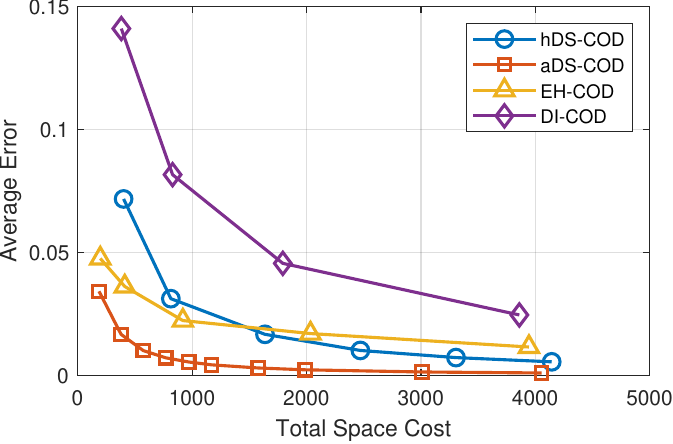}}\hfill
    
\vspace{-0.2cm}

    \subfigure[SYNTHETIC(1)]{\includegraphics[width=0.25\textwidth]{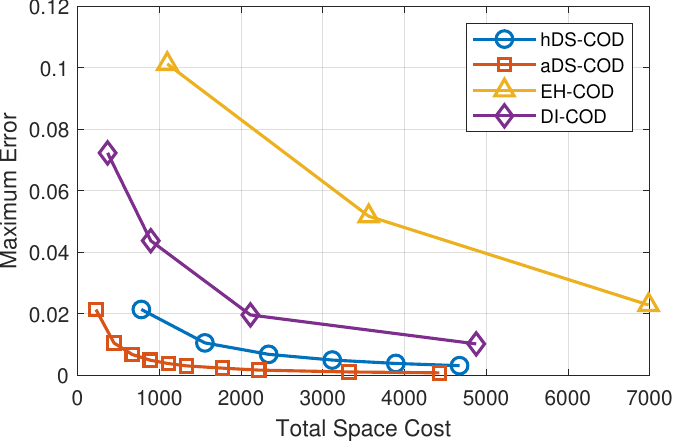}}\hfill
    \subfigure[SYNTHETIC(2)]{\includegraphics[width=0.25\textwidth]{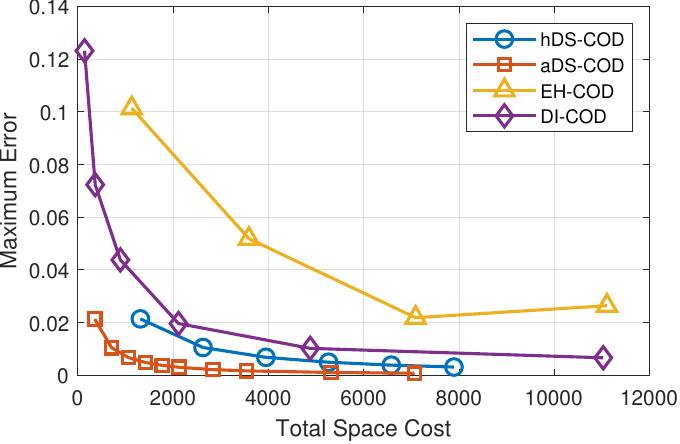}}\hfill
    \subfigure[APR]{\includegraphics[width=0.25\textwidth]{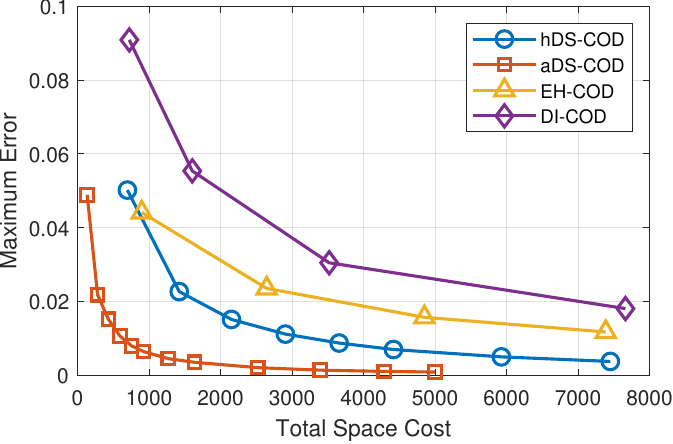}}\hfill
    \subfigure[PAN11]{\includegraphics[width=0.25\textwidth]{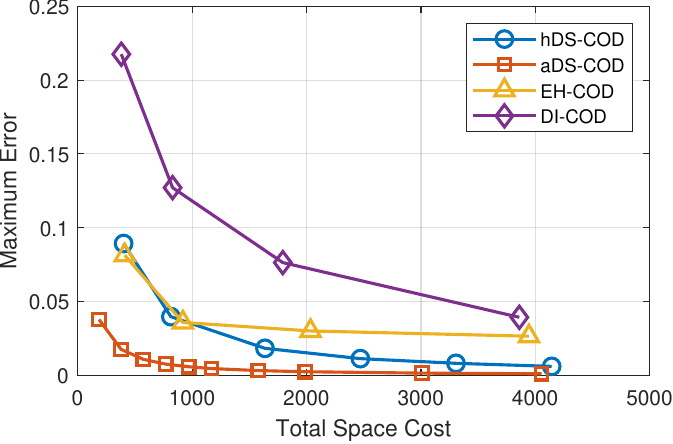}}\hfill
    
\vspace{-0.3cm}

    \caption{The plot of correlation error against total space cost.}
    \label{fig:error vs space}
    
\vspace{0.5cm}

    \subfigure[average error vs. final sketch size]{\includegraphics[width=0.25\textwidth]{fig/euro_1,1.pdf}}\hfill
    \subfigure[maximum error vs. final sketch size]{\includegraphics[width=0.25\textwidth]{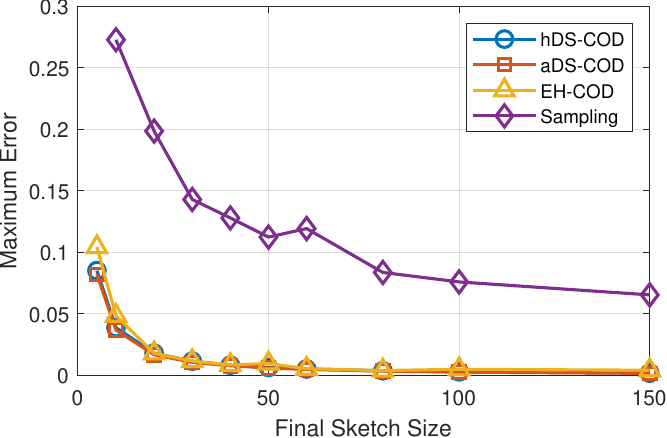}}\hfill
    \subfigure[average error vs. total space cost]{\includegraphics[width=0.25\textwidth]{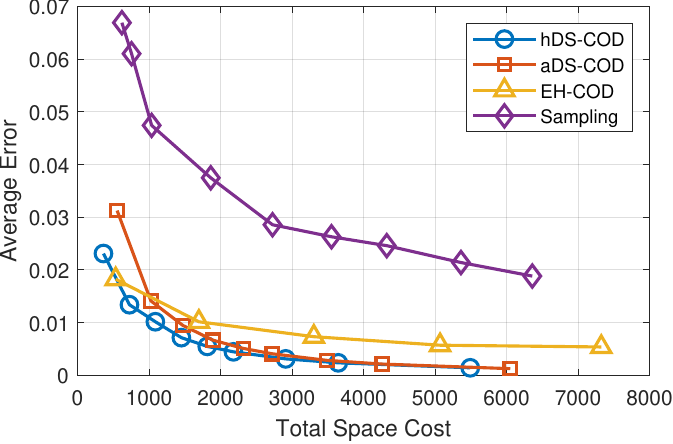}}\hfill
    \subfigure[maximum error vs. total space cost]{\includegraphics[width=0.25\textwidth]{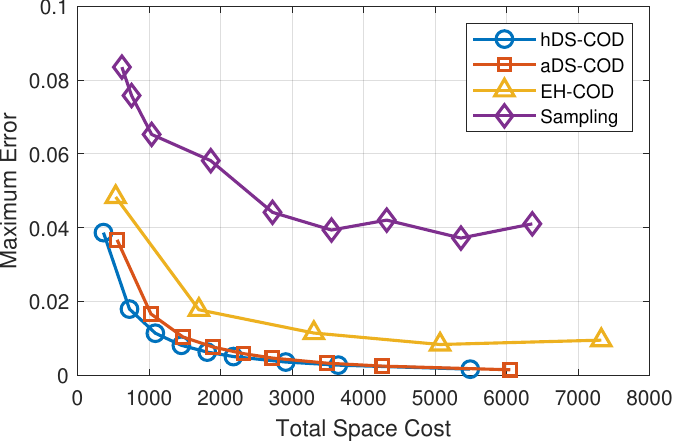}}\hfill

\vspace{-0.3cm}
    \caption{The plot of error-space trade-off on EURO dataset (Time-based Window).}
    \label{fig:time-based}
\end{figure*}

\paragraph{Performance} Figure \ref{fig:error vs l} and Figure \ref{fig:error vs space} illustrate the space efficiency comparison of the algorithms on sequence-based datasets. Panels (a-d) show the average errors across all sampled windows, while panels (e-h) display the maximum errors.

Figure \ref{fig:error vs l} evaluates the compression effect of the final sketch. The hDS-COD, aDS-COD, and EH-COD show similar compression performances. But the DS series is more stable, particularly on the synthetic datasets, where they significantly outperform EH-COD and DI-COD. The performance of hDS-COD and aDS-COD is nearly the same, indicating that the adaptive threshold trick in aDS-COD does not have a noticeable negative impact on it, maintaining the same error as hDS-COD.

Figure \ref{fig:error vs space} measures the total space cost of the algorithms. hDS-COD and aDS-COD show a significant advantage over existing methods, as they can achieve the  $\varepsilon$-approximation error with much less space. For the same space cost, the correlation errors of hDS-COD and aDS-COD are much smaller than those of EH-COD and DI-COD. Also, aDS-COD has better space efficiency than hDS-COD because aDS only uses a single-level structure while hDS requires $\log R+1$ levels. We find that hDS-COD requires more space on  SYNTHETIC(2) dataset compared to SYNTHETIC(1) dataset. This phenomenon occurs because SYNTHETIC(2) dataset has a larger $R$, which confirms the dependence on $R$ as stated in Theorem~\ref{thm:hds}. 

Figure \ref{fig:time-based} compares the performance of algorithms on time-based windows. Panels (a) and (b) present the error against the final sketch size, which show that our aDS-COD and hDS-COD algorithms enjoy similar performance as EH-COD and significantly outperform the sampling algorithm. On the other hand, as shown in panels (c) and (d), our methods outperform baselines in terms of total space cost.

\section{Conclusion}
In this paper, we introduced a deterministic algorithm called DS-COD for AMM over sliding windows, providing rigorous error bounds and complexity analysis. For both the sequence-based and time-based models, we demonstrated that DS-COD achieves an optimal space-error tradeoff, matching the lower bounds we established. Additionally, we present an improved version of DS-COD with an adaptive adjustment mechanism, which shows excellent empirical performance. Experiments on synthetic and real-world datasets validate the efficiency of our algorithms.
Notice that ~\citet{yao2024approximate} also studied sliding window AMM with sparse data. It would be an interesting future work to extend our methods to the sparse case.

\FloatBarrier

\begin{acks}
\end{acks}

\bibliographystyle{ACM-Reference-Format}
\balance
\bibliography{references}


\begin{thebibliography}{33}


\ifx \showCODEN    \undefined \def \showCODEN     #1{\unskip}     \fi
\ifx \showDOI      \undefined \def \showDOI       #1{#1}\fi
\ifx \showISBNx    \undefined \def \showISBNx     #1{\unskip}     \fi
\ifx \showISBNxiii \undefined \def \showISBNxiii  #1{\unskip}     \fi
\ifx \showISSN     \undefined \def \showISSN      #1{\unskip}     \fi
\ifx \showLCCN     \undefined \def \showLCCN      #1{\unskip}     \fi
\ifx \shownote     \undefined \def \shownote      #1{#1}          \fi
\ifx \showarticletitle \undefined \def \showarticletitle #1{#1}   \fi
\ifx \showURL      \undefined \def \showURL       {\relax}        \fi
\providecommand\bibfield[2]{#2}
\providecommand\bibinfo[2]{#2}
\providecommand\natexlab[1]{#1}
\providecommand\showeprint[2][]{arXiv:#2}

\bibitem[Abdi and Williams(2010)]%
        {abdi2010principal}
\bibfield{author}{\bibinfo{person}{Herv{\'e} Abdi} {and} \bibinfo{person}{Lynne~J Williams}.} \bibinfo{year}{2010}\natexlab{}.
\newblock \showarticletitle{Principal component analysis}.
\newblock \bibinfo{journal}{\emph{Wiley interdisciplinary reviews: computational statistics}} \bibinfo{volume}{2}, \bibinfo{number}{4} (\bibinfo{year}{2010}), \bibinfo{pages}{433--459}.
\newblock


\bibitem[Agarwal et~al\mbox{.}(2019)]%
        {agarwal2019efficient}
\bibfield{author}{\bibinfo{person}{Naman Agarwal}, \bibinfo{person}{Brian Bullins}, \bibinfo{person}{Xinyi Chen}, \bibinfo{person}{Elad Hazan}, \bibinfo{person}{Karan Singh}, \bibinfo{person}{Cyril Zhang}, {and} \bibinfo{person}{Yi Zhang}.} \bibinfo{year}{2019}\natexlab{}.
\newblock \showarticletitle{Efficient full-matrix adaptive regularization}. In \bibinfo{booktitle}{\emph{International Conference on Machine Learning}}. PMLR, \bibinfo{pages}{102--110}.
\newblock


\bibitem[Arasu and Manku(2004)]%
        {arasu2004approximate}
\bibfield{author}{\bibinfo{person}{Arvind Arasu} {and} \bibinfo{person}{Gurmeet~Singh Manku}.} \bibinfo{year}{2004}\natexlab{}.
\newblock \showarticletitle{Approximate counts and quantiles over sliding windows}. In \bibinfo{booktitle}{\emph{Proceedings of the twenty-third ACM SIGMOD-SIGACT-SIGART symposium on Principles of database systems}}. \bibinfo{pages}{286--296}.
\newblock


\bibitem[Becker et~al\mbox{.}(2011)]%
        {becker2011beyond}
\bibfield{author}{\bibinfo{person}{Hila Becker}, \bibinfo{person}{Mor Naaman}, {and} \bibinfo{person}{Luis Gravano}.} \bibinfo{year}{2011}\natexlab{}.
\newblock \showarticletitle{Beyond trending topics: Real-world event identification on twitter}. In \bibinfo{booktitle}{\emph{Proceedings of the international AAAI conference on web and social media}}, Vol.~\bibinfo{volume}{5}. \bibinfo{pages}{438--441}.
\newblock


\bibitem[Blalock and Guttag(2021)]%
        {blalock2021multiplying}
\bibfield{author}{\bibinfo{person}{Davis Blalock} {and} \bibinfo{person}{John Guttag}.} \bibinfo{year}{2021}\natexlab{}.
\newblock \showarticletitle{Multiplying matrices without multiplying}. In \bibinfo{booktitle}{\emph{International Conference on Machine Learning}}. PMLR, \bibinfo{pages}{992--1004}.
\newblock


\bibitem[Braverman et~al\mbox{.}(2020)]%
        {braverman2020near}
\bibfield{author}{\bibinfo{person}{Vladimir Braverman}, \bibinfo{person}{Petros Drineas}, \bibinfo{person}{Cameron Musco}, \bibinfo{person}{Christopher Musco}, \bibinfo{person}{Jalaj Upadhyay}, \bibinfo{person}{David~P Woodruff}, {and} \bibinfo{person}{Samson Zhou}.} \bibinfo{year}{2020}\natexlab{}.
\newblock \showarticletitle{Near optimal linear algebra in the online and sliding window models}. In \bibinfo{booktitle}{\emph{2020 IEEE 61st Annual Symposium on Foundations of Computer Science (FOCS)}}. IEEE, \bibinfo{pages}{517--528}.
\newblock


\bibitem[Cohen and Lewis(1999)]%
        {cohen1999approximating}
\bibfield{author}{\bibinfo{person}{Edith Cohen} {and} \bibinfo{person}{David~D Lewis}.} \bibinfo{year}{1999}\natexlab{}.
\newblock \showarticletitle{Approximating matrix multiplication for pattern recognition tasks}.
\newblock \bibinfo{journal}{\emph{Journal of Algorithms}} \bibinfo{volume}{30}, \bibinfo{number}{2} (\bibinfo{year}{1999}), \bibinfo{pages}{211--252}.
\newblock


\bibitem[Cohen et~al\mbox{.}(2015a)]%
        {cohen2015dimensionality}
\bibfield{author}{\bibinfo{person}{Michael~B Cohen}, \bibinfo{person}{Sam Elder}, \bibinfo{person}{Cameron Musco}, \bibinfo{person}{Christopher Musco}, {and} \bibinfo{person}{Madalina Persu}.} \bibinfo{year}{2015}\natexlab{a}.
\newblock \showarticletitle{Dimensionality reduction for k-means clustering and low rank approximation}. In \bibinfo{booktitle}{\emph{Proceedings of the forty-seventh annual ACM symposium on Theory of computing}}. \bibinfo{pages}{163--172}.
\newblock


\bibitem[Cohen et~al\mbox{.}(2015b)]%
        {cohen2015optimal}
\bibfield{author}{\bibinfo{person}{Michael~B Cohen}, \bibinfo{person}{Jelani Nelson}, {and} \bibinfo{person}{David~P Woodruff}.} \bibinfo{year}{2015}\natexlab{b}.
\newblock \showarticletitle{Optimal approximate matrix product in terms of stable rank}.
\newblock \bibinfo{journal}{\emph{arXiv preprint arXiv:1507.02268}} (\bibinfo{year}{2015}).
\newblock


\bibitem[Datar et~al\mbox{.}(2002)]%
        {datar2002maintaining}
\bibfield{author}{\bibinfo{person}{Mayur Datar}, \bibinfo{person}{Aristides Gionis}, \bibinfo{person}{Piotr Indyk}, {and} \bibinfo{person}{Rajeev Motwani}.} \bibinfo{year}{2002}\natexlab{}.
\newblock \showarticletitle{Maintaining stream statistics over sliding windows}.
\newblock \bibinfo{journal}{\emph{SIAM journal on computing}} \bibinfo{volume}{31}, \bibinfo{number}{6} (\bibinfo{year}{2002}), \bibinfo{pages}{1794--1813}.
\newblock


\bibitem[Dhillon(2001)]%
        {dhillon2001co}
\bibfield{author}{\bibinfo{person}{Inderjit~S Dhillon}.} \bibinfo{year}{2001}\natexlab{}.
\newblock \showarticletitle{Co-clustering documents and words using bipartite spectral graph partitioning}. In \bibinfo{booktitle}{\emph{Proceedings of the 7th ACM SIGKDD international conference on Knowledge discovery and data mining}}. \bibinfo{pages}{269--274}.
\newblock


\bibitem[Drineas et~al\mbox{.}(2006)]%
        {drineas2006fast}
\bibfield{author}{\bibinfo{person}{Petros Drineas}, \bibinfo{person}{Ravi Kannan}, {and} \bibinfo{person}{Michael~W Mahoney}.} \bibinfo{year}{2006}\natexlab{}.
\newblock \showarticletitle{Fast Monte Carlo algorithms for matrices I: Approximating matrix multiplication}.
\newblock \bibinfo{journal}{\emph{SIAM J. Comput.}} \bibinfo{volume}{36}, \bibinfo{number}{1} (\bibinfo{year}{2006}), \bibinfo{pages}{132--157}.
\newblock


\bibitem[Duchi et~al\mbox{.}(2011)]%
        {duchi2011adaptive}
\bibfield{author}{\bibinfo{person}{John Duchi}, \bibinfo{person}{Elad Hazan}, {and} \bibinfo{person}{Yoram Singer}.} \bibinfo{year}{2011}\natexlab{}.
\newblock \showarticletitle{Adaptive subgradient methods for online learning and stochastic optimization.}
\newblock \bibinfo{journal}{\emph{The Journal of Machine Learning Research}} \bibinfo{volume}{12}, \bibinfo{number}{7} (\bibinfo{year}{2011}).
\newblock


\bibitem[Francis and Raimond(2022)]%
        {francis2022practical}
\bibfield{author}{\bibinfo{person}{Deena~P Francis} {and} \bibinfo{person}{Kumudha Raimond}.} \bibinfo{year}{2022}\natexlab{}.
\newblock \showarticletitle{A practical streaming approximate matrix multiplication algorithm}.
\newblock \bibinfo{journal}{\emph{Journal of King Saud University-Computer and Information Sciences}} \bibinfo{volume}{34}, \bibinfo{number}{1} (\bibinfo{year}{2022}), \bibinfo{pages}{1455--1465}.
\newblock


\bibitem[Greenacre et~al\mbox{.}(2022)]%
        {greenacre2022principal}
\bibfield{author}{\bibinfo{person}{Michael Greenacre}, \bibinfo{person}{Patrick~JF Groenen}, \bibinfo{person}{Trevor Hastie}, \bibinfo{person}{Alfonso~Iodice d’Enza}, \bibinfo{person}{Angelos Markos}, {and} \bibinfo{person}{Elena Tuzhilina}.} \bibinfo{year}{2022}\natexlab{}.
\newblock \showarticletitle{Principal component analysis}.
\newblock \bibinfo{journal}{\emph{Nature Reviews Methods Primers}} \bibinfo{volume}{2}, \bibinfo{number}{1} (\bibinfo{year}{2022}), \bibinfo{pages}{100}.
\newblock


\bibitem[Gupta et~al\mbox{.}(2018)]%
        {gupta2018oversketch}
\bibfield{author}{\bibinfo{person}{Vipul Gupta}, \bibinfo{person}{Shusen Wang}, \bibinfo{person}{Thomas Courtade}, {and} \bibinfo{person}{Kannan Ramchandran}.} \bibinfo{year}{2018}\natexlab{}.
\newblock \showarticletitle{Oversketch: Approximate matrix multiplication for the cloud}. In \bibinfo{booktitle}{\emph{2018 IEEE International Conference on Big Data (Big Data)}}. IEEE, \bibinfo{pages}{298--304}.
\newblock


\bibitem[Hasan and Abdulazeez(2021)]%
        {hasan2021review}
\bibfield{author}{\bibinfo{person}{Basna Mohammed~Salih Hasan} {and} \bibinfo{person}{Adnan~Mohsin Abdulazeez}.} \bibinfo{year}{2021}\natexlab{}.
\newblock \showarticletitle{A review of principal component analysis algorithm for dimensionality reduction}.
\newblock \bibinfo{journal}{\emph{Journal of Soft Computing and Data Mining}} \bibinfo{volume}{2}, \bibinfo{number}{1} (\bibinfo{year}{2021}), \bibinfo{pages}{20--30}.
\newblock


\bibitem[Joshi and Hadi(2015)]%
        {joshi2015review}
\bibfield{author}{\bibinfo{person}{Manish Joshi} {and} \bibinfo{person}{Theyazn~Hassn Hadi}.} \bibinfo{year}{2015}\natexlab{}.
\newblock \showarticletitle{A review of network traffic analysis and prediction techniques}.
\newblock \bibinfo{journal}{\emph{arXiv preprint arXiv:1507.05722}} (\bibinfo{year}{2015}).
\newblock


\bibitem[Kyrillidis et~al\mbox{.}(2014)]%
        {kyrillidis2014approximate}
\bibfield{author}{\bibinfo{person}{Anastasios Kyrillidis}, \bibinfo{person}{Michail Vlachos}, {and} \bibinfo{person}{Anastasios Zouzias}.} \bibinfo{year}{2014}\natexlab{}.
\newblock \showarticletitle{Approximate matrix multiplication with application to linear embeddings}. In \bibinfo{booktitle}{\emph{2014 IEEE International Symposium on Information Theory}}. Ieee, \bibinfo{pages}{2182--2186}.
\newblock


\bibitem[Lee and Ting(2006)]%
        {lee2006simpler}
\bibfield{author}{\bibinfo{person}{Lap-Kei Lee} {and} \bibinfo{person}{HF Ting}.} \bibinfo{year}{2006}\natexlab{}.
\newblock \showarticletitle{A simpler and more efficient deterministic scheme for finding frequent items over sliding windows}. In \bibinfo{booktitle}{\emph{Proceedings of the twenty-fifth ACM SIGMOD-SIGACT-SIGART symposium on Principles of database systems}}. \bibinfo{pages}{290--297}.
\newblock


\bibitem[Luo et~al\mbox{.}(2021)]%
        {luo2021revisiting}
\bibfield{author}{\bibinfo{person}{Luo Luo}, \bibinfo{person}{Cheng Chen}, \bibinfo{person}{Guangzeng Xie}, {and} \bibinfo{person}{Haishan Ye}.} \bibinfo{year}{2021}\natexlab{}.
\newblock \showarticletitle{Revisiting Co-Occurring Directions: Sharper Analysis and Efficient Algorithm for Sparse Matrices}. In \bibinfo{booktitle}{\emph{Proceedings of the AAAI Conference on Artificial Intelligence}}, Vol.~\bibinfo{volume}{35}. \bibinfo{pages}{8793--8800}.
\newblock


\bibitem[Luo et~al\mbox{.}(2019)]%
        {luo2019robust}
\bibfield{author}{\bibinfo{person}{Luo Luo}, \bibinfo{person}{Cheng Chen}, \bibinfo{person}{Zhihua Zhang}, \bibinfo{person}{Wu-Jun Li}, {and} \bibinfo{person}{Tong Zhang}.} \bibinfo{year}{2019}\natexlab{}.
\newblock \showarticletitle{Robust frequent directions with application in online learning}.
\newblock \bibinfo{journal}{\emph{The Journal of Machine Learning Research}} \bibinfo{volume}{20}, \bibinfo{number}{1} (\bibinfo{year}{2019}).
\newblock


\bibitem[Mroueh et~al\mbox{.}(2017)]%
        {mroueh2017co}
\bibfield{author}{\bibinfo{person}{Youssef Mroueh}, \bibinfo{person}{Etienne Marcheret}, {and} \bibinfo{person}{Vaibahava Goel}.} \bibinfo{year}{2017}\natexlab{}.
\newblock \showarticletitle{Co-occurring directions sketching for approximate matrix multiply}. In \bibinfo{booktitle}{\emph{Artificial Intelligence and Statistics}}. PMLR, \bibinfo{pages}{567--575}.
\newblock


\bibitem[Naseem et~al\mbox{.}(2010)]%
        {naseem2010linear}
\bibfield{author}{\bibinfo{person}{Imran Naseem}, \bibinfo{person}{Roberto Togneri}, {and} \bibinfo{person}{Mohammed Bennamoun}.} \bibinfo{year}{2010}\natexlab{}.
\newblock \showarticletitle{Linear regression for face recognition}.
\newblock \bibinfo{journal}{\emph{IEEE transactions on pattern analysis and machine intelligence (TPAMI)}} \bibinfo{volume}{32}, \bibinfo{number}{11} (\bibinfo{year}{2010}), \bibinfo{pages}{2106--2112}.
\newblock


\bibitem[Plancher et~al\mbox{.}(2019)]%
        {plancher2019application}
\bibfield{author}{\bibinfo{person}{Brian Plancher}, \bibinfo{person}{Camelia~D Brumar}, \bibinfo{person}{Iulian Brumar}, \bibinfo{person}{Lillian Pentecost}, \bibinfo{person}{Saketh Rama}, {and} \bibinfo{person}{David Brooks}.} \bibinfo{year}{2019}\natexlab{}.
\newblock \showarticletitle{Application of approximate matrix multiplication to neural networks and distributed SLAM}. In \bibinfo{booktitle}{\emph{2019 IEEE High Performance Extreme Computing Conference (HPEC)}}. IEEE, \bibinfo{pages}{1--7}.
\newblock


\bibitem[Sarlos(2006)]%
        {sarlos2006improved}
\bibfield{author}{\bibinfo{person}{Tamas Sarlos}.} \bibinfo{year}{2006}\natexlab{}.
\newblock \showarticletitle{Improved approximation algorithms for large matrices via random projections}. In \bibinfo{booktitle}{\emph{2006 IEEE 47th Annual Symposium on Foundations of Computer Science (FOCS)}}. IEEE, \bibinfo{pages}{143--152}.
\newblock


\bibitem[Wan and Zhang(2021)]%
        {wan2021approximate}
\bibfield{author}{\bibinfo{person}{Yuanyu Wan} {and} \bibinfo{person}{Lijun Zhang}.} \bibinfo{year}{2021}\natexlab{}.
\newblock \showarticletitle{Approximate Multiplication of Sparse Matrices with Limited Space}. In \bibinfo{booktitle}{\emph{Proceedings of the AAAI Conference on Artificial Intelligence}}, Vol.~\bibinfo{volume}{35}. \bibinfo{pages}{10058--10066}.
\newblock


\bibitem[Wan and Zhang(2022)]%
        {TPAMI2022Wan}
\bibfield{author}{\bibinfo{person}{Yuanyu Wan} {and} \bibinfo{person}{Lijun Zhang}.} \bibinfo{year}{2022}\natexlab{}.
\newblock \showarticletitle{Efficient Adaptive Online Learning via Frequent Directions}.
\newblock \bibinfo{journal}{\emph{IEEE Transactions on Pattern Analysis and Machine Intelligence (TPAMI)}} \bibinfo{volume}{44}, \bibinfo{number}{10} (\bibinfo{year}{2022}), \bibinfo{pages}{6910--6923}.
\newblock


\bibitem[Wei et~al\mbox{.}(2016)]%
        {wei2016matrix}
\bibfield{author}{\bibinfo{person}{Zhewei Wei}, \bibinfo{person}{Xuancheng Liu}, \bibinfo{person}{Feifei Li}, \bibinfo{person}{Shuo Shang}, \bibinfo{person}{Xiaoyong Du}, {and} \bibinfo{person}{Ji-Rong Wen}.} \bibinfo{year}{2016}\natexlab{}.
\newblock \showarticletitle{Matrix sketching over sliding windows}. In \bibinfo{booktitle}{\emph{Proceedings of the 2016 International Conference on Management of Data}}. \bibinfo{pages}{1465--1480}.
\newblock


\bibitem[Woodruff et~al\mbox{.}(2014)]%
        {woodruff2014sketching}
\bibfield{author}{\bibinfo{person}{David~P Woodruff} {et~al\mbox{.}}} \bibinfo{year}{2014}\natexlab{}.
\newblock \showarticletitle{Sketching as a tool for numerical linear algebra}.
\newblock \bibinfo{journal}{\emph{Foundations and Trends{\textregistered} in Theoretical Computer Science}} \bibinfo{volume}{10}, \bibinfo{number}{1--2} (\bibinfo{year}{2014}), \bibinfo{pages}{1--157}.
\newblock


\bibitem[Yao et~al\mbox{.}(2024)]%
        {yao2024approximate}
\bibfield{author}{\bibinfo{person}{Ziqi Yao}, \bibinfo{person}{Lianzhi Li}, \bibinfo{person}{Mingsong Chen}, \bibinfo{person}{Xian Wei}, {and} \bibinfo{person}{Cheng Chen}.} \bibinfo{year}{2024}\natexlab{}.
\newblock \showarticletitle{Approximate Matrix Multiplication over Sliding Windows}. In \bibinfo{booktitle}{\emph{Proceedings of the 30th ACM SIGKDD Conference on Knowledge Discovery and Data Mining}}. \bibinfo{pages}{3896--3906}.
\newblock


\bibitem[Ye et~al\mbox{.}(2016)]%
        {ye2016frequent}
\bibfield{author}{\bibinfo{person}{Qiaomin Ye}, \bibinfo{person}{Luo Luo}, {and} \bibinfo{person}{Zhihua Zhang}.} \bibinfo{year}{2016}\natexlab{}.
\newblock \showarticletitle{Frequent direction algorithms for approximate matrix multiplication with applications in CCA}.
\newblock \bibinfo{journal}{\emph{Computational Complexity}} \bibinfo{volume}{1}, \bibinfo{number}{m3} (\bibinfo{year}{2016}), \bibinfo{pages}{2}.
\newblock


\bibitem[Yin et~al\mbox{.}(2024)]%
        {yin2024optimal}
\bibfield{author}{\bibinfo{person}{Hanyan Yin}, \bibinfo{person}{Dongxie Wen}, \bibinfo{person}{Jiajun Li}, \bibinfo{person}{Zhewei Wei}, \bibinfo{person}{Xiao Zhang}, \bibinfo{person}{Zengfeng Huang}, {and} \bibinfo{person}{Feifei Li}.} \bibinfo{year}{2024}\natexlab{}.
\newblock \showarticletitle{Optimal Matrix Sketching over Sliding Windows}.
\newblock \bibinfo{journal}{\emph{arXiv preprint arXiv:2405.07792}} (\bibinfo{year}{2024}).
\newblock


\end{thebibliography}


\begin{thebibliography}{10}
\providecommand{\url}[1]{#1}
\csname url@samestyle\endcsname
\providecommand{\newblock}{\relax}
\providecommand{\bibinfo}[2]{#2}
\providecommand{\BIBentrySTDinterwordspacing}{\spaceskip=0pt\relax}
\providecommand{\BIBentryALTinterwordstretchfactor}{4}
\providecommand{\BIBentryALTinterwordspacing}{\spaceskip=\fontdimen2\font plus
\BIBentryALTinterwordstretchfactor\fontdimen3\font minus \fontdimen4\font\relax}
\providecommand{\BIBforeignlanguage}[2]{{%
\expandafter\ifx\csname l@#1\endcsname\relax
\typeout{** WARNING: IEEEtran.bst: No hyphenation pattern has been}%
\typeout{** loaded for the language `#1'. Using the pattern for}%
\typeout{** the default language instead.}%
\else
\language=\csname l@#1\endcsname
\fi
#2}}
\providecommand{\BIBdecl}{\relax}
\BIBdecl

\bibitem{abdi2010principal}
H.~Abdi and L.~J. Williams, ``Principal component analysis,'' \emph{Wiley interdisciplinary reviews: computational statistics}, vol.~2, no.~4, pp. 433--459, 2010.

\bibitem{mroueh2017co}
Y.~Mroueh, E.~Marcheret, and V.~Goel, ``Co-occurring directions sketching for approximate matrix multiply,'' in \emph{Artificial Intelligence and Statistics}.\hskip 1em plus 0.5em minus 0.4em\relax PMLR, 2017, pp. 567--575.

\bibitem{yao2024approximate}
Z.~Yao, L.~Li, M.~Chen, X.~Wei, and C.~Chen, ``Approximate matrix multiplication over sliding windows,'' in \emph{Proceedings of the 30th ACM SIGKDD Conference on Knowledge Discovery and Data Mining}, 2024, pp. 3896--3906.

\bibitem{luo2021revisiting}
L.~Luo, C.~Chen, G.~Xie, and H.~Ye, ``Revisiting co-occurring directions: Sharper analysis and efficient algorithm for sparse matrices,'' in \emph{Proceedings of the AAAI Conference on Artificial Intelligence}, vol.~35, no.~10, 2021, pp. 8793--8800.

\bibitem{wan2021approximate}
Y.~Wan and L.~Zhang, ``Approximate multiplication of sparse matrices with limited space,'' in \emph{Proceedings of the AAAI Conference on Artificial Intelligence}, vol.~35, no.~11, 2021, pp. 10\,058--10\,066.

\bibitem{hasan2021review}
B.~M.~S. Hasan and A.~M. Abdulazeez, ``A review of principal component analysis algorithm for dimensionality reduction,'' \emph{Journal of Soft Computing and Data Mining}, vol.~2, no.~1, pp. 20--30, 2021.

\bibitem{blalock2021multiplying}
D.~Blalock and J.~Guttag, ``Multiplying matrices without multiplying,'' in \emph{International Conference on Machine Learning}.\hskip 1em plus 0.5em minus 0.4em\relax PMLR, 2021, pp. 992--1004.

\bibitem{cohen1999approximating}
E.~Cohen and D.~D. Lewis, ``Approximating matrix multiplication for pattern recognition tasks,'' \emph{Journal of Algorithms}, vol.~30, no.~2, pp. 211--252, 1999.

\bibitem{francis2022practical}
D.~P. Francis and K.~Raimond, ``A practical streaming approximate matrix multiplication algorithm,'' \emph{Journal of King Saud University-Computer and Information Sciences}, vol.~34, no.~1, pp. 1455--1465, 2022.

\bibitem{osawa2017accelerating}
K.~Osawa, A.~Sekiya, H.~Naganuma, and R.~Yokota, ``Accelerating matrix multiplication in deep learning by using low-rank approximation,'' in \emph{2017 International Conference on High Performance Computing \& Simulation (HPCS)}.\hskip 1em plus 0.5em minus 0.4em\relax IEEE, 2017, pp. 186--192.

\bibitem{gupta2018oversketch}
V.~Gupta, S.~Wang, T.~Courtade, and K.~Ramchandran, ``Oversketch: Approximate matrix multiplication for the cloud,'' in \emph{2018 IEEE International Conference on Big Data (Big Data)}.\hskip 1em plus 0.5em minus 0.4em\relax IEEE, 2018, pp. 298--304.

\bibitem{greenacre2022principal}
M.~Greenacre, P.~J. Groenen, T.~Hastie, A.~I. d’Enza, A.~Markos, and E.~Tuzhilina, ``Principal component analysis,'' \emph{Nature Reviews Methods Primers}, vol.~2, no.~1, p. 100, 2022.

\bibitem{datar2002maintaining}
M.~Datar, A.~Gionis, P.~Indyk, and R.~Motwani, ``Maintaining stream statistics over sliding windows,'' \emph{SIAM journal on computing}, vol.~31, no.~6, pp. 1794--1813, 2002.

\bibitem{arasu2004approximate}
A.~Arasu and G.~S. Manku, ``Approximate counts and quantiles over sliding windows,'' in \emph{Proceedings of the twenty-third ACM SIGMOD-SIGACT-SIGART symposium on Principles of database systems}, 2004, pp. 286--296.

\bibitem{lee2006simpler}
L.-K. Lee and H.~Ting, ``A simpler and more efficient deterministic scheme for finding frequent items over sliding windows,'' in \emph{Proceedings of the twenty-fifth ACM SIGMOD-SIGACT-SIGART symposium on Principles of database systems}, 2006, pp. 290--297.

\bibitem{becker2011beyond}
H.~Becker, M.~Naaman, and L.~Gravano, ``Beyond trending topics: Real-world event identification on twitter,'' in \emph{Proceedings of the international AAAI conference on web and social media}, vol.~5, no.~1, 2011, pp. 438--441.

\bibitem{braverman2020near}
V.~Braverman, P.~Drineas, C.~Musco, C.~Musco, J.~Upadhyay, D.~P. Woodruff, and S.~Zhou, ``Near optimal linear algebra in the online and sliding window models,'' in \emph{2020 IEEE 61st Annual Symposium on Foundations of Computer Science (FOCS)}.\hskip 1em plus 0.5em minus 0.4em\relax IEEE, 2020, pp. 517--528.

\bibitem{joshi2015review}
M.~Joshi and T.~H. Hadi, ``A review of network traffic analysis and prediction techniques,'' \emph{arXiv preprint arXiv:1507.05722}, 2015.

\bibitem{wei2016matrix}
Z.~Wei, X.~Liu, F.~Li, S.~Shang, X.~Du, and J.-R. Wen, ``Matrix sketching over sliding windows,'' in \emph{Proceedings of the 2016 International Conference on Management of Data}, 2016, pp. 1465--1480.

\bibitem{yin2024optimal}
H.~Yin, D.~Wen, J.~Li, Z.~Wei, X.~Zhang, Z.~Huang, and F.~Li, ``Optimal matrix sketching over sliding windows,'' \emph{arXiv preprint arXiv:2405.07792}, 2024.

\bibitem{drineas2006fast}
P.~Drineas, R.~Kannan, and M.~W. Mahoney, ``Fast monte carlo algorithms for matrices i: Approximating matrix multiplication,'' \emph{SIAM Journal on Computing}, vol.~36, no.~1, pp. 132--157, 2006.

\bibitem{sarlos2006improved}
T.~Sarlos, ``Improved approximation algorithms for large matrices via random projections,'' in \emph{2006 IEEE 47th Annual Symposium on Foundations of Computer Science (FOCS)}.\hskip 1em plus 0.5em minus 0.4em\relax IEEE, 2006, pp. 143--152.

\bibitem{magen2011low}
A.~Magen and A.~Zouzias, ``Low rank matrix-valued chernoff bounds and approximate matrix multiplication,'' in \emph{Proceedings of the twenty-second annual ACM-SIAM symposium on Discrete Algorithms}.\hskip 1em plus 0.5em minus 0.4em\relax SIAM, 2011, pp. 1422--1436.

\bibitem{cohen2015optimal}
M.~B. Cohen, J.~Nelson, and D.~P. Woodruff, ``Optimal approximate matrix product in terms of stable rank,'' \emph{arXiv preprint arXiv:1507.02268}, 2015.

\bibitem{ye2016frequent}
Q.~Ye, L.~Luo, and Z.~Zhang, ``Frequent direction algorithms for approximate matrix multiplication with applications in cca,'' \emph{Computational Complexity}, vol.~1, no.~m3, p.~2, 2016.

\bibitem{ghashami2016frequent}
M.~Ghashami, E.~Liberty, J.~M. Phillips, and D.~P. Woodruff, ``Frequent directions: Simple and deterministic matrix sketching,'' \emph{SIAM Journal on Computing}, vol.~45, no.~5, pp. 1762--1792, 2016.

\bibitem{datar2016sliding}
M.~Datar and R.~Motwani, ``The sliding-window computation model and results,'' in \emph{Data Stream Management: Processing High-Speed Data Streams}.\hskip 1em plus 0.5em minus 0.4em\relax Springer, 2016, pp. 149--165.

\bibitem{woodruff2014sketching}
D.~P. Woodruff \emph{et~al.}, ``Sketching as a tool for numerical linear algebra,'' \emph{Foundations and Trends{\textregistered} in Theoretical Computer Science}, vol.~10, no. 1--2, pp. 1--157, 2014.

\bibitem{naseem2010linear}
I.~Naseem, R.~Togneri, and M.~Bennamoun, ``Linear regression for face recognition,'' \emph{IEEE transactions on pattern analysis and machine intelligence (TPAMI)}, vol.~32, no.~11, pp. 2106--2112, 2010.

\bibitem{kyrillidis2014approximate}
A.~Kyrillidis, M.~Vlachos, and A.~Zouzias, ``Approximate matrix multiplication with application to linear embeddings,'' in \emph{2014 IEEE International Symposium on Information Theory}.\hskip 1em plus 0.5em minus 0.4em\relax Ieee, 2014, pp. 2182--2186.

\bibitem{cohen2015dimensionality}
M.~B. Cohen, S.~Elder, C.~Musco, C.~Musco, and M.~Persu, ``Dimensionality reduction for k-means clustering and low rank approximation,'' in \emph{Proceedings of the forty-seventh annual ACM symposium on Theory of computing}, 2015, pp. 163--172.

\bibitem{dhillon2001co}
I.~S. Dhillon, ``Co-clustering documents and words using bipartite spectral graph partitioning,'' in \emph{Proceedings of the 7th ACM SIGKDD international conference on Knowledge discovery and data mining}, 2001, pp. 269--274.

\bibitem{agarwal2019efficient}
N.~Agarwal, B.~Bullins, X.~Chen, E.~Hazan, K.~Singh, C.~Zhang, and Y.~Zhang, ``Efficient full-matrix adaptive regularization,'' in \emph{International Conference on Machine Learning}.\hskip 1em plus 0.5em minus 0.4em\relax PMLR, 2019, pp. 102--110.

\bibitem{duchi2011adaptive}
J.~Duchi, E.~Hazan, and Y.~Singer, ``Adaptive subgradient methods for online learning and stochastic optimization.'' \emph{The Journal of Machine Learning Research}, vol.~12, no.~7, 2011.

\bibitem{luo2019robust}
L.~Luo, C.~Chen, Z.~Zhang, W.-J. Li, and T.~Zhang, ``Robust frequent directions with application in online learning,'' \emph{The Journal of Machine Learning Research}, vol.~20, no.~1, 2019.

\bibitem{TPAMI2022Wan}
Y.~Wan and L.~Zhang, ``Efficient adaptive online learning via frequent directions,'' \emph{IEEE Transactions on Pattern Analysis and Machine Intelligence (TPAMI)}, vol.~44, no.~10, pp. 6910--6923, 2022.

\bibitem{hotelling1936relations}
H.~Hotelling, ``Relations between two sets of variates,'' \emph{Biometrika}, 1936.

\bibitem{babcock2002models}
B.~Babcock, S.~Babu, M.~Datar, R.~Motwani, and J.~Widom, ``Models and issues in data stream systems,'' in \emph{Proceedings of the twenty-first ACM SIGMOD-SIGACT-SIGART symposium on Principles of database systems}, 2002, pp. 1--16.

\bibitem{plancher2019application}
B.~Plancher, C.~D. Brumar, I.~Brumar, L.~Pentecost, S.~Rama, and D.~Brooks, ``Application of approximate matrix multiplication to neural networks and distributed slam,'' in \emph{2019 IEEE High Performance Extreme Computing Conference (HPEC)}.\hskip 1em plus 0.5em minus 0.4em\relax IEEE, 2019, pp. 1--7.

\bibitem{manku2002approximate}
G.~S. Manku and R.~Motwani, ``Approximate frequency counts over data streams,'' in \emph{VLDB'02: Proceedings of the 28th International Conference on Very Large Databases}.\hskip 1em plus 0.5em minus 0.4em\relax Elsevier, 2002, pp. 346--357.

\bibitem{papapetrou2015sketching}
O.~Papapetrou, M.~Garofalakis, and A.~Deligiannakis, ``Sketching distributed sliding-window data streams,'' \emph{The VLDB Journal}, vol.~24, pp. 345--368, 2015.

\bibitem{borassi2020sliding}
M.~Borassi, A.~Epasto, S.~Lattanzi, S.~Vassilvitskii, and M.~Zadimoghaddam, ``Sliding window algorithms for k-clustering problems,'' \emph{Advances in Neural Information Processing Systems}, vol.~33, 2020.

\bibitem{musco2015randomized}
C.~Musco and C.~Musco, ``Randomized block krylov methods for stronger and faster approximate singular value decomposition,'' \emph{Advances in Neural Information Processing Systems}, vol.~28, 2015.

\bibitem{zhang2017tracking}
H.~Zhang, Z.~Huang, Z.~Wei, W.~Zhang, and X.~Lin, ``Tracking matrix approximation over distributed sliding windows,'' in \emph{2017 IEEE 33rd International Conference on Data Engineering (ICDE)}.\hskip 1em plus 0.5em minus 0.4em\relax IEEE, 2017, pp. 833--844.

\bibitem{babcock2002sampling}
B.~Babcock, M.~Datar, and R.~Motwani, ``Sampling from a moving window over streaming data,'' in \emph{Proceedings of the thirteenth annual ACM-SIAM symposium on Discrete algorithms}.\hskip 1em plus 0.5em minus 0.4em\relax SIAM, 2002, pp. 633--634.

\bibitem{longbo2007priority}
Z.~Longbo, L.~Zhanhuai, Z.~Yiqiang, Y.~Min, and Z.~Yang, ``A priority random sampling algorithm for time-based sliding windows over weighted streaming data,'' in \emph{Proceedings of the 2007 ACM symposium on Applied computing}, 2007, pp. 453--456.

\bibitem{badeau2004sliding}
R.~Badeau, G.~Richard, and B.~David, ``Sliding window adaptive svd algorithms,'' \emph{IEEE Transactions on Signal Processing}, vol.~52, no.~1, pp. 1--10, 2004.

\bibitem{liberty2013simple}
E.~Liberty, ``Simple and deterministic matrix sketching,'' in \emph{Proceedings of the 19th ACM SIGKDD international conference on Knowledge discovery and data mining}, 2013, pp. 581--588.

\bibitem{prettenhofer2010cross}
P.~Prettenhofer and B.~Stein, ``Cross-language text classification using structural correspondence learning,'' in \emph{Proceedings of the 48th annual meeting of the association for computational linguistics}, 2010, pp. 1118--1127.

\bibitem{potthast2011cross}
M.~Potthast, A.~Barr{\'o}n-Cedeno, B.~Stein, and P.~Rosso, ``Cross-language plagiarism detection,'' \emph{Language Resources and Evaluation}, vol.~45, pp. 45--62, 2011.

\bibitem{potthast2010evaluation}
M.~Potthast, B.~Stein, A.~Barr{\'o}n-Cede{\~n}o, and P.~Rosso, ``An evaluation framework for plagiarism detection,'' in \emph{Coling 2010: Posters}, 2010, pp. 997--1005.

\bibitem{koehn2005europarl}
P.~Koehn, ``Europarl: A parallel corpus for statistical machine translation,'' in \emph{Proceedings of machine translation summit x: papers}, 2005, pp. 79--86.

\bibitem{ferrero2016multilingual}
J.~Ferrero, F.~Agnes, L.~Besacier, and D.~Schwab, ``A multilingual, multi-style and multi-granularity dataset for cross-language textual similarity detection,'' in \emph{Proceedings of the Tenth International Conference on Language Resources and Evaluation (LREC)}, 2016, pp. 4162--4169.

\end{thebibliography}

\appendix
\onecolumn
\section{Proof of Lemma \ref{lem:remove}} \label{apdx:remove}
\begin{proof}
    Let $\mR$ denotes $\mR_x\mR_y^\top=\mU\mathbf{\Sigma}\mV^\top$, then the original product is $\mA\mB^\top = \mQ_x \mR \mQ_y^\top$. When the first snapshot dumped, the change in $\mR$ can be expressed as: $\overline{\mR} = \mR-\sigma_1\vu_1\vv_1^\top$. The product after this dump is $\overline{\mA\mB^\top}=\mQ_x\overline{\mR}\mQ_y^\top$. Now, define $\overline{\mR}_x = \mR_x-\vu_1\vu_1^\top\mR_x,\overline{\mR}_y = \mR_y-\vv_1\vv_1^\top\mR_y$. We can prove $\overline{\mR}_x\overline{\mR}_y^\top=\overline{\mR}$ as below,
   \begin{align*}
         \overline{\mR}_x\overline{\mR}_y^\top &= (\mI-\vu_1\vu_1^\top)\mR_x\mR_y^\top(\mI-\vv_1\vv_1^\top)  \\
         &= \mR-\vu_1\vu_1^\top\mR-\mR\vv_1\vv_1^\top+\vu_1\vu_1^\top\mR\vv_1\vv_1^\top   \\
         &= \mR-\sigma_1\vu_1\vv_1^\top 
    \end{align*}
    where the last equality holds due to $\vu_1 \vu_1^\top \mR = \vu_1 \vu_1^\top \sum_{i} \sigma_i \vu_i \vv_i^\top = \sigma_1 \vu_1 \vv_1^\top$, and similarly for the terms $\mR\vv_1\vv_1^\top$ and $\vu_1\vu_1^\top\mR\vv_1\vv_1^\top$. Therefore, $\overline{\mA\mB^\top}=\mQ_x\overline{\mR}_x\overline{\mR}_y^\top\mQ_y^\top$. Now we can reconstruct $\overline{\mA} = \mQ_x\overline{\mR}_x$ and $\overline{\mB} = \mQ_y\overline{\mR}_y$. Apparently, they are equivalent to $\mC$ and $\mD$ with the first column removed. 
\end{proof}

\section{Proof of Theorem \ref{thm:hds}}\label{apdx:hds}
\begin{proof}
    
Let $\mX_{T-N,T}\mY_{T-N,T}^\top$ be the original product within the window $[T-N,T]$. The algorithm \ref{alg:hdscod} returns an approximation $\mA_{T-N,T}\mB_{T-N,T}^\top+\hat{\mC}_{(k-1)N+1,T}\hat{\mD}_{(k-1)N+1,T}^\top$, where $\mA_{T-N,T}\mB_{T-N,T}^\top$ denotes the key component of the product core consisting of snapshots, and $\hat{\mC}_{(k-1)N+1,T}\hat{\mD}_{(k-1)N+1,T}^\top$ represents the sketch residual product.  Here, $k=\operatorname*{max}(1,\lfloor(T-1)/N\rfloor)$ and $(k-1)N+1$ represents the last timestamp of the N-restart. Define $\mC_{T-N,T}\mD_{T-N,T}^\top = \mX_{T-N,T}\mY_{T-N,N}^\top-\mA_{T-N,T}\mB_{T-N,T}^\top$ to be the real residual product. For simplicity, we denote $\hat{\mC}_{(k-1)N+1,T}$ as $\hat{\mC}_T$, and similarly for the other terms. The correlation error can be expressed as:
\begin{align}
    &\quad \|    \mX_{T-N,T}\mY_{T-N,T}^\top-(\mA_{T-N,T}\mB_{T-N,T}^\top+\hat{\mC}_{(k-1)N+1,T}\hat{\mD}_{(k-1)N+1,T}^\top)    \|_2 \nonumber\\
    &= \|     \mX_T\mY_T^\top - \mX_{T-N}\mY_{T-N}^\top - \mA_T\mB_T^\top +\mA_{T-N}\mB_{T-N}^\top  -  \hat{\mC}_T\hat{\mD}_T^\top     \|_2 \nonumber\\
    &= \|     (\mC_T\mD_T^\top-\hat{\mC}_T\hat{\mD}_T^\top) - (\mC_{T-N}\mD_{T-N}^\top - \hat{\mC}_{T-N}\hat{\mD}_{T-N}^\top) +   \hat{\mC}_{T-N}\hat{\mD}_{T-N}^\top   \|_2 \nonumber\\
    &\leq \| \mC_T\mD_T^\top-\hat{\mC}_T\hat{\mD}_T^\top \|_2 + \| \mC_{T-N}\mD_{T-N}^\top - \hat{\mC}_{T-N}\hat{\mD}_{T-N}^\top \|_2 + \| \hat{\mC}_{T-N}\hat{\mD}_{T-N}^\top \|_2
\end{align}
Define the iterative error $\Delta_t = \vx_t\vy_t^\top + \hat{\mC}_{t-1}\hat{\mD}_{t-1}^\top - (\hat{\mC}_{t}\hat{\mD}_{t}^\top + \va_t \vb_t^\top)$. Sum it up from $(k-1)N+1$ to $T$: 
\[
    \sum_{t=(k-1)N+1}^\top{\Delta_t}  = \mX_T\mY_T^\top - \hat{\mC}_{T}\hat{\mD}_{T}^\top - \mA_T\mB_T^\top = \mC_T\mD_T^\top - \hat{\mC}_{T}\hat{\mD}_{T}^\top.
\]
By the definition of COD, we have
\begin{align}
    \Delta_t &= \mQ_x^{(t)}\mU_t\mathbf{\Sigma}_t\mV_t^\top\mQ_y^{(t)T} - \mQ_x^{(t)}\mU_t \cdot \max{(\mathbf{\Sigma_t}-\sigma_{\ell}^{(t)} \mI,0)} \cdot \mV_t^\top\mQ_y^{(t)\top} \nonumber\\
    &= \mQ_x^{(t)} \mU_t \cdot \min{(\mathbf{\Sigma_t},\sigma_{\ell}^{(t)} \mI)} \cdot \mV_t^\top\mQ_y^{(t)\top} \nonumber
\end{align}
Thus, 
\[
\| \mC_T\mD_T^\top - \hat{\mC}_{T}\hat{\mD}_{T}^\top \|_2 \leq \sum_{t=(k-1)N+1}^\top \| \Delta_t \|_2  = \sum_{t=(k-1)N+1}^\top \sigma_\ell^{(t)}.
\]
According to the proof of theorem 2 of \cite{mroueh2017co}, we can deduce that $\sum_{t=(k-1)N+1}^\top \sigma_\ell^{(t)} \leq \frac{2}{\ell}\|\mX_T\|_F\|\mY_T\|_F $.

Suppose the maximum squared norm of column within the window is $2^i, i\in[0,\log R]$, we have
\begin{gather}
    \| \mC_T\mD_T^\top-\hat{\mC}_T\hat{\mD}_T^\top \|_2 \leq 2\varepsilon\|\mX_T\|_F\|\mY_T\|_F  \leq 2\varepsilon \cdot2^i(T-(k-1)N) \nonumber \\
    \| \mC_{T-N}\mD_{T-N}^\top - \hat{\mC}_{T-N}\hat{\mD}_{T-N}^\top \|_2 \leq 2\varepsilon\|\mX_{T-N}\|_F\|\mY_{T-N}\|_F  \leq 2\varepsilon \cdot2^i(T-N-(k-1)N) \nonumber \\
    \| \hat{\mC}_{T-N}\hat{\mD}_{T-N}^\top\| \leq 2\varepsilon \cdot2^iN \nonumber.
\end{gather}

Therefore, Equation (2) can be bounded by
\[
    4\varepsilon \cdot 2^i(T-(k-1)N) \leq 2^{i+3}\varepsilon N.
\]
To satisfy error $2^{i+3}\varepsilon N \leq \alpha\varepsilon \|\mX_{T-N,T}\|_F\|\mY_{T-N,T}\|_F$, we choose the level $i$:
\[
\log_2{\frac{\alpha\|\mX_{T-N,T}\|_F\|\mY_{T-N,T}\|_F}{8N}}-1 \leq i \leq \log_2{\frac{\alpha\|\mX_{T-N,T}\|_F\|\mY_{T-N,T}\|_F}{8N}}
\]
, where $i$ always exist when $\alpha\geq8$.
\end{proof}

\section{Complexity analysis of Algorithm \ref{alg:hdscod}} \label{apdx:comlexity}
\paragraph{Space Complexity.} Each DS-COD sketch contains a snapshot queue $S$ and a pair of residual matrices $\mA$ and $\mB$. In the hierarchical structure, each snapshot queues is bounded by $O(\ell)$ column vectors, and each residual matrix contains at most $O(\ell)$ columns. Since there are $\log R+1$ levels, the total space complexity is $O\left(\frac{m_x + m_y}{\varepsilon} (\log R+1)\right)$.
\paragraph{Time Complexity.} The time complexity of the DS-COD algorithm is mainly driven by its update operations. In case 1, the amortized cost of each CS operation, performed every $\ell$ columns, is $O((m_x + m_y) \ell)$. The rank-1 update operations for maintaining the covariance matrices $\mK_A$ and $\mK_B$ cost $O((m_x + m_y) \ell)$, while LDL and SVD decompositions each require $O(\ell^3)$.
Lines 16-23 of the algorithm may produce multiple snapshots, but the total number of snapshots in any window is bounded by the window size $N$, so this cost can be amortized over each update. The cost of computing the snapshots $\va_j$ and $\vb_j$ is $O((m_x + m_y) \ell)$. The updates to the matrices $\overline{\mA}$ and $\overline{\mB}$, which are used to remove the influence of the snapshots, take the following form:
$$\overline{\mA} = \mA - \frac{1}{\sigma_1} \mA (\mR_y^\top \vv_1)(\vu_1^\top \mR_x), \quad \overline{\mB} = \mB - \frac{1}{\sigma_1} \mB (\mR_x^\top \vu_1)(\vv_1^\top \mR_y),$$
which requires $O((m_x + m_y) \ell)$.
Next, the update to the covariance matrix $\overline{\mK}_A$ is calculated as:
\begin{align}
    \overline{\mK}_A &=\mK_A-\mK_A\mP_A-\mP_A^\top\mK_A+\mP_A^\top\mK_A\mP_A \nonumber\\
    &= \mK_A-(\mA^\top\mA\mP_A+\mP_A^\top\mA^\top\mA-\mP_A^\top\mA^\top\mA\mP_A) \nonumber\\
    &= \mK_A-(\mA^\top\mA\mP_A+\mP_A^\top\mA^\top\overline{\mA}) \nonumber\\
    &= \mK_A-(\mA^\top\mA-\mA^\top\overline{\mA}+\mP_A^\top\mA^\top\overline{\mA}) \nonumber\\
    &= \mA^\top\overline{\mA} - \mP_A^\top\mA^\top\overline{\mA} \nonumber&
\end{align}
Where the term $\mA^\top \overline{\mA}$ is computed as:
$$\mA^\top \overline{\mA} = \mA^\top (\mA - \mA \mP_A) = \mK_A - \mK_A \mP_A = \mK_A - \frac{1}{\sigma_1} \mK_A (\mR_y^\top \vv_1)(\vu_1^\top \mR_x),$$
which costs $O(\ell^2)$. Similarly, the term $\mP_A^\top \mA^\top \overline{\mA}$ requires $O(\ell^2)$ as well, since:
$$\mP_A^\top \mA^\top \overline{\mA} = \frac{1}{\sigma_1} \mR_x^\top \vu_1 \left( (\vv_1^\top \mR_y) (\mA^\top \overline{\mA}) \right).$$
Thus, the amortized time complexity for a single update step in Algorithm \ref{alg:update} is $O((m_x + m_y)\ell + \ell^3)$. Assuming $m_x + m_y = \Omega(\ell^2)$, the update cost simplifies to $O((m_x + m_y)\ell)$. Given that there are $\log{R}+1$ levels to update, the total cost of DS-COD is $O\left( \frac{m_x + m_y}{\varepsilon} (\log{R}+1) \right)$.

\section{Proof of Theorem \ref{thm:lowerbound}}\label{apdx:lowerbound}
\begin{proof}
Without loss of generality, we suppose that $m_x\le m_y$ and let $\ell=1/\varepsilon$. By Lemma \ref{lem:matrix_set}, we can construct a set of matrix pairs $\fZ_{\ell/4}$ of cardinality $2^{\Omega((m_x+m_y)\ell)}$, where (i) each pair $(\mX^{(i)},\mY^{(i)})$ satisfies $\mathbf{X}^{(i)}\in\mathbb{R}^{m_x \times \frac{\ell}{4}}$,  $\mathbf{Y}^{(i)}\in\mathbb{R}^{m_y \times \frac{\ell}{4}}$ and $\mathbf{X}^{(i)\top} \mathbf{X}^{(i)} = \mathbf{Y}^{(i)\top} \mathbf{Y}^{(i)} = \mathbf{I}_{\ell/4}$; (ii) $\left\|\mathbf{X}^{(i)} \mathbf{Y}^{(i)\top} - \mathbf{X}^{(j)} \mathbf{Y}^{(j)\top}\right\|_2 > \frac{1}{2}$ for any $i\ne j$. We can select $\log{R} + 1$ matrix pairs from the set $\mathcal{Z}_{\ell/4}$ with repetition, arrange them in a sequence and label them from right to left as $0,1,2,\dots,\log{R}$, making total number of distinct arrangements is $S=2^{\Omega((m_x+m_y)\ell(\log R+1))}$. We call the matrix pair labeled $i$ as block $i$. Then, we adjust these blocks to construct a worst-case instance through the following steps:
\begin{enumerate}
    \item For each block $i$, we scale their unit column vectors by a factor of $\sqrt{\frac{2^i N}{\ell}}$ and thus the size of block $i$ satisfies $\|\mX_{i}\|_F\|\mY_{i}\|_F = \frac{2^i N}{4}$.
    
    \item For blocks where $i > \log_2\left(\frac{\ell R}{N}\right)$, to ensure that all column pairs  satisfies $1\le\|\vx\|\|\vy\|\le R$. Thus we increase the number of columns of these blocks from $\frac{\ell}{4}$ to $\frac{\ell}{4} \cdot 2^{i - \log_2 \frac{\ell R}{N}}$ by duplicate the matrix for $2^{i - \log_2 \frac{\ell R}{N}}$ times and scale the squared norm of each column to $R$. 
    Consequently, the total number of columns can be computed as 
        \[
        \frac{\ell}{4} \cdot \left(\log_2{\frac{\ell R}{N}}+1\right) + \frac{\ell}{4}\cdot\sum_{i=\log\frac{\ell R}{N}+1}^{\log_2 R} 2^{i - \log_2 \frac{\ell R}{N}}= \frac{\ell}{4} \left(\log_2{\frac{\ell R}{N}}+1\right)+\frac{\ell}{2}\left( 2^{\log_2{R} - \log_2{\left(\frac{\ell R}{N}\right)}} - 1 \right) = \frac{\ell}{4} \log{\frac{\ell R}{2N}} + \frac{N}{2}.
        \]
    Since we have assumed that $N \geq \frac{1}{2\varepsilon} \log{\frac{R}{\varepsilon}}$ and $\ell=\frac{1}{\varepsilon}$, then we have $\frac{\ell}{4} \log{\frac{\ell R}{2N}} + \frac{N}{2}\le N$, which means a window can contain all selected blocks.
    
    \item All columns outside the aforementioned blocks, including those that have not yet arrived, are set to zero vectors. Additionally, we append an extra dimension to each vector, setting its value to $1$. 
\end{enumerate}

We define $(\mX_W^i,\mY_W^i)$ as the active data over the sliding window of length $N$ at the moment when $i+1,i+2,\dots,\log{R}$ blocks have expired. Suppose a sliding window algorithm provides $\frac{\varepsilon}{9}$-correlation sketch $(\mA_W^i,\mB_W^i)$ and $(\mA_W^{i-1},\mB_W^{i-1})$ for $(\mX_W^i,\mY_W^i)$ and $(\mX_W^{i-1},\mY_W^{i-1})$, respectively. The guarantee of the algorithm indicates
\begin{align}
    \| \mX_W^i{\mY_W^{i}}^\top - \mA_W^i{\mB_W^{i}}^\top \|_2 &\leq \frac{\varepsilon}{9} \|\mX_W^i\|_F\|\mY_W^i\|_F \leq  \frac{\varepsilon}{9} \sum_{j=0}^i{\|\mX^{(j)}\|_F\|\mY^{(j)}\|_F} = \frac{\varepsilon}{9}\left( \frac{N}{4}\cdot2^{i+1}+\frac{3N}{4} \right)\nonumber
\end{align}
Similarly, we can obtain the following result:  
\[
\| \mX_W^{i-1} {\mY_W^{i-1}}^\top - \mA_W^{i-1} {\mB_W^{i-1}}^\top \|_2 \leq \frac{\varepsilon}{9} n \left( \frac{N}{4} \cdot 2^i + \frac{3N}{4} \right).
\]
Therefore, we can approximate block $i$ with $\mA_i\mB_i^\top=\mA_W^i{\mB_W^{i}}^\top-\mA_W^{i-1} {\mB_W^{i-1}}^\top$. Then we have
\begin{align}
    &\| \mX_i{\mY_{i}}^\top - \mA_{i}{\mB_{i}}^\top \|_2 \nonumber\\
    = &\| (\mX_W^i{\mY_W^{i}}^\top- \mX_W^{i-1}{\mY_W^{i-1}}^\top) - (\mA^i{\mB^{i}}^\top - \mA^{i-1}{\mB^{i-1}}^\top) \|_2 \nonumber\\
    \leq & \| \mX_W^i{\mY_W^{i}}^\top- \mA^i{\mB^{i}}^\top\|_2 + \| \mX_W^{i-1}{\mY_W^{i-1}}^\top - \mA^{i-1}{\mB^{i-1}}^\top \|_2 \nonumber\\
    \leq &\frac{\varepsilon}{9} \left( \frac{3}{4}N\cdot2^i+\frac{3}{2}N \right) \nonumber\\
    \leq &  \frac{1}{\ell} \cdot\frac{2^i N}{4} \nonumber
\end{align}

Notice that if we use two different matrix pairs $(\mathbf{X}^{(i)}, \mathbf{Y}^{(i)})$ and $(\mathbf{X}^{(j)}, \mathbf{Y}^{(j)})$ from $\fZ_{\ell/4}$ to construct block $i$, we always have 
$$\left\|\mathbf{X}^{(i)} \mathbf{Y}^{(i)\top} - \mathbf{X}^{(j)} \mathbf{Y}^{(j)\top}\right\|_2 > \frac{1}{2}\cdot \frac{2^i N}{\ell}.$$ 
Hence the algorithm can encode a sequence of $\log R+1$ blocks. Since we can have $S=2^{\Omega((m_x+m_y)\ell(\log R+1))}$ sequence in total, which means the lower bound of space complexity is $\log S=\Omega((m_x+m_y)\ell(\log R+1))$.

\end{proof}

\section{Proof of Theorem \ref{thm:lowerbound_time}}\label{apdx:lowerbound_time}

\begin{proof}
    Similar to the proof of Theorem \ref{thm:lowerbound}, we first construct a set of matrix pairs \( \mathcal{Z}_{\ell/4} \). We then select \( \log{(NR/\ell)} +1\) matrix pairs with repetition, arrange them in a sequence, and label them from right to left as \( 0, 1, \dots, \log{(NR/\ell)} \), resulting in \( S = 2^{\Omega((m_x + m_y)\ell \log{(NR/\ell)})} \) distinct arrangements. Finally, we construct a worst-case instance using the following steps:
    \begin{enumerate}
        \item For each block \( i \), we scale the unit column vectors by \( \sqrt{2^i} \), so the size of block \( i \) satisfies \( \|\mX_{i}\|_F \|\mY_{i}\|_F = \frac{2^i \ell}{4} \).
        
        \item For blocks where \( i > \log_2{R} \), we increase the number of columns from \( \frac{\ell}{4} \) to \( \frac{\ell}{4} \cdot 2^{i - \log_2{R}} \) to ensure \( 1 \leq \|\mathbf{x}\| \|\mathbf{y}\| \leq R \). The total number of columns, \( \frac{\ell}{4} \log{\frac{N}{2}} \), should be bounded by \( N \), leading to \( N \geq \frac{\ell}{2} \log{\frac{R}{2}} \).
        
        \item Columns outside the mentioned blocks, including those not yet arrived, are set to zero vectors.
    \end{enumerate}

    We define \( (\mX_W^i, \mY_W^i) \) as the active data in the sliding window of length \( N \) after \( i+1, i+2, \dots, \log{(NR/\ell)} \) blocks have expired. Suppose a sliding window algorithm provides a \( \frac{\varepsilon}{3} \)-correlation sketch \( (\mA_W^i, \mB_W^i) \) and \( (\mA_W^{i-1}, \mB_W^{i-1}) \) for \( (\mX_W^i, \mY_W^i) \) and \( (\mX_W^{i-1}, \mY_W^{i-1}) \), respectively. The algorithm's guarantee indicates:
    \begin{align}
        \| \mX_W^i{\mY_W^{i}}^\top - \mA_W^i{\mB_W^{i}}^\top \|_2 &\leq \frac{\varepsilon}{3} \|\mX_W^i\|_F\|\mY_W^i\|_F \nonumber\\
        &\leq  \frac{\varepsilon}{3} \sum_{j=0}^i{\|\mX^{(j)}\|_F\|\mY^{(j)}\|_F} \nonumber\\
        &= \frac{\varepsilon}{3}\left( \frac{\ell}{4}\cdot2^{i+1}-\frac{\ell}{4} \right)\nonumber
    \end{align}
    Similarly, we obtain:
    \[
    \| \mX_W^{i-1} {\mY_W^{i-1}}^\top - \mA_W^{i-1} {\mB_W^{i-1}}^\top \|_2 \leq \frac{\varepsilon}{3}  \left( \frac{\ell}{4} \cdot 2^i - \frac{\ell}{4} \right).
    \]
    Therefore, we approximate block \( i \) as \( \mA_i \mB_i^\top = \mA_W^i{\mB_W^{i}}^\top - \mA_W^{i-1} {\mB_W^{i-1}}^\top \). Then we have:
    \begin{align}
        &\| \mX_i {\mY_{i}}^\top - \mA_{i}{\mB_{i}}^\top \|_2 \nonumber\\
        = &\| (\mX_W^i {\mY_W^{i}}^\top - \mX_W^{i-1} {\mY_W^{i-1}}^\top) - (\mA^i {\mB^{i}}^\top - \mA^{i-1} {\mB^{i-1}}^\top) \|_2 \nonumber\\
        \leq & \| \mX_W^i {\mY_W^{i}}^\top - \mA^i {\mB^{i}}^\top \|_2 + \| \mX_W^{i-1} {\mY_W^{i-1}}^\top - \mA^{i-1} {\mB^{i-1}}^\top \|_2 \nonumber\\
        \leq & \frac{\varepsilon}{3} \left( \frac{3}{4}\ell\cdot2^i - \frac{1}{2}\ell \right) \nonumber\\
        \leq &  \frac{1}{\ell} \cdot \frac{2^i \ell}{4} \nonumber
    \end{align}
    
    Notice that when using two different matrix pairs \( (\mathbf{X}^{(i)}, \mathbf{Y}^{(i)}) \) and \( (\mathbf{X}^{(j)}, \mathbf{Y}^{(j)}) \) from \( \mathcal{Z}_{\ell/4} \) to construct block \( i \), we always have:
    $$\left\|\mathbf{X}^{(i)} \mathbf{Y}^{(i)\top} - \mathbf{X}^{(j)} \mathbf{Y}^{(j)\top}\right\|_2 > \frac{1}{2} \cdot 2^i.$$ 
    Hence, the algorithm can encode a sequence of \( \log{(NR/\ell)} +1 \) blocks. Since the total number of distinct sequences is \( S = 2^{\Omega((m_x+m_y)\ell \log{(NR/\ell)})} \), the lower bound for the space complexity is \( \log{S} = \Omega((m_x + m_y)\ell \log{(NR/\ell)}) \).
\end{proof}

\end{document}